\pdfoutput=1

\documentclass[11pt,twoside,a4paper,cmspaper,final,collab]{cms-tdr}
\def\svnVersion{364699}\def\cmsCernNoTag{CERN-PH-EP/2015-122}\def\cmsCernDate{\today}\def\cmsMessage{Published in the European Physical Journal C as \href{http://dx.doi.org/10.1140/epjc/s10052-016-4219-1}{\doi{10.1140/epjc/s10052-016-4219-1.}}}

\begin{document}\cmsNoteHeader{SMP-14-016}

\hyphenation{had-ron-i-za-tion}
\hyphenation{cal-or-i-me-ter}
\hyphenation{de-vices}
\RCS$HeadURL: svn+ssh://svn.cern.ch/reps/tdr2/papers/SMP-14-016/trunk/SMP-14-016.tex $
\RCS$Id: SMP-14-016.tex 347901 2016-06-17 22:08:49Z alverson $

\newlength\cmsFigWidth
\ifthenelse{\boolean{cms@external}}{\setlength\cmsFigWidth{0.85\columnwidth}}{\setlength\cmsFigWidth{0.4\textwidth}}
\ifthenelse{\boolean{cms@external}}{\providecommand{\cmsLeft}{top\xspace}}{\providecommand{\cmsLeft}{left\xspace}}
\ifthenelse{\boolean{cms@external}}{\providecommand{\cmsRight}{bottom\xspace}}{\providecommand{\cmsRight}{right\xspace}}
\providecommand{\PGmmp}{\ensuremath{\mu^\mp}\xspace} 
\providecommand{\experr}{\ensuremath{\,\text{(exp)}}\xspace}
\providecommand{\WW}{\ensuremath{\PWp\PWm}\xspace}
\providecommand{\WZ}{\ensuremath{\PW\Z}\xspace}
\providecommand{\ZZ}{\ensuremath{\Z\Z}\xspace}
\providecommand{\NA}{---}
\ifthenelse{\boolean{cms@external}}{\providecommand{\breakhere}{\linebreak[4]}}{\providecommand{\breakhere}{\relax}}
\newcommand{\ppww}{\ensuremath{\Pp\Pp\to \PWp\PWm}\xspace}
\newcommand{\V}{\ensuremath{\mathrm{V}}\xspace}
\newcommand{\dyll}{\ensuremath{\Z/\gamma^*\to \ell^+\ell^-}\xspace}
\newcommand{\dymm}{\ensuremath{\Z/\gamma^*\to \PGmp\PGmm}\xspace}
\newcommand{\dytt}{\ensuremath{\Z/\gamma^* \to\PGt^+\PGt^-}\xspace}
\newcommand{\mll}{\ensuremath{m_{\ell\ell}}\xspace}
\newcommand{\MT}{\ensuremath{m_{\mathrm{T}}}\xspace}
\newcommand{\rootSHat}{\ensuremath{\sqrt{\hat{s}}\xspace}}
\newcommand{\SHat}{\ensuremath{\hat{s}\xspace}}

\newcolumntype{x}{D{,}{\,\pm\,}{-1}}

\cmsNoteHeader{SMP-14-016}
\title{Measurement of the $\PWp\PWm$ cross section in pp collisions at $\sqrt{s} = 8$\TeV and limits on anomalous gauge couplings}
\titlerunning{$\PWp\PWm$ cross sections and ATGC limits at 8\TeV}

\date{\today}

\abstract{
A measurement of the W boson pair production cross section in proton-proton collisions at
$\sqrt{s} = 8$\TeV is presented.
The data collected with the CMS detector at the LHC correspond to an
integrated luminosity of 19.4\fbinv. The
$\PWp\PWm$ candidates are
selected from events with two charged leptons, electrons or muons, and large
missing transverse energy. The
measured $\PWp\PWm$ cross section is
$60.1\pm 0.9\stat\pm 3.2\experr\pm 3.1\thy\pm 1.6\lum\unit{pb} =
60.1\pm 4.8\unit{pb}$, consistent with the standard model prediction.
The $\PWp\PWm$ cross sections are also measured in two different
fiducial phase space regions. The normalized differential cross section is measured as a function
of kinematic variables of the final-state charged leptons and compared with several
perturbative QCD predictions. Limits on anomalous gauge couplings associated with dimension-six
operators are also given in the framework of an effective field theory. The corresponding 95\%
confidence level intervals are
$-5.7 <  c_{\mathrm{WWW}}/\Lambda^2  < 5.9\TeV^{-2}$,
$-11.4 <  c_{\mathrm{W}}/\Lambda^2  < 5.4\TeV^{-2}$,
$-29.2 <  c_{\mathrm{B}}/\Lambda^2  < 23.9\TeV^{-2}$, in the HISZ basis.
}

\hypersetup{%
pdfauthor={CMS Collaboration},%
pdftitle={Measurement of the W+ W- cross section in pp collisions at sqrt(s) = 8 TeV and limits on anomalous gauge couplings},%
pdfsubject={CMS},%
pdfkeywords={CMS, physics, WW cross section, ATGC}}

\maketitle

\section{Introduction}\label{sec:intro}

{\tolerance=800
The standard model (SM) description of electroweak and strong interactions
can be tested through precision measurements of the $\WW$ production cross
section at hadron colliders. Among the massive vector boson pair production processes,
$\WW$ has the largest cross section.

At the CERN LHC, the SM vector boson pair production is dominated by the $s$-channel and $t$-channel
quark-antiquark ($\qqbar$) annihilation diagrams, while the gluon-gluon ($\cPg\cPg$) diagrams contribute only 3\%
to the total production cross section~\cite{MCFM}. Previous cross section results on $\WW$ production in pp collisions at a
center-of-mass energy of $\sqrt{s} = 7\TeV$ are reported to be $52.4\pm 2.0\stat\pm 4.5\syst\pm 1.62\lum\unit{pb}$
by CMS~\cite{PublishedWW7tev} and $54.4\pm 4.0\stat\pm 3.9\syst\pm 2.0\lum\unit{pb}$ by ATLAS~\cite{ATLASPublishedWW7tev}. Results at $\sqrt{s} = 8\TeV$ are reported by CMS using 3.5\fbinv of data~\cite{PublishedWW} with a measured value of
$69.9\pm 2.8\stat\pm 5.6\syst\pm 3.1\lum\unit{pb}$. Also, a cross section measurement of
$\WW$ production in $\Pp\Pap$ collisions at $\sqrt{s} = 1.96$\TeV has been recently reported by CDF to be
$14.0\pm 0.6\stat{^{+1.2}_{-1.0}}\syst\pm 0.8\lum\unit{pb}$~\cite{CDFxsec}.
Next-to-next-to-leading-order (NNLO) calculations for the $\WW$ production in pp collisions at
$\sqrt{s} = 8\TeV$ predict a cross section of
$\sigma^{\mathrm{NNLO}} (\Pp\Pp \to \WW) = 59.8^{+1.3}_{-1.1}\unit{pb}$~\cite{NNLOXsec}. In this $\WW$ production calculation, processes
involving the SM Higgs boson are not considered; it is estimated they would increase the total cross section by
about 8\% for the Higgs boson mass of 125\GeV~\cite{HiggsXS}.

We measure the $\WW$ production cross section in the fully leptonic decay channel by selecting events
with two high transverse momentum (\pt) electrons or muons ($\Pep\Pem$, $\PGmp\PGmm$, $\Pepm\PGmmp$),
large missing transverse energy (\MET), and zero or one jet with high \pt. We provide a more precise
measurement than previous results~\cite{PublishedWW} by using an improved analysis strategy and a larger data sample.
The \pt of the $\WW$ system receives large higher-order corrections because of the restriction on the number
of jets. The dominant $\qqbar$ component of the signal production is modeled by resumming
the large higher-order corrections to the $\WW$ \pt distribution, thus improving the signal efficiency
determination~\cite{ptresum,ptresum3}. The expected contribution, based on simulation, from Higgs-boson-mediated
processes to the observed signal yield is subtracted. The data correspond to a total accumulated luminosity
of 19.4\fbinv at $\sqrt{s} = 8\TeV$.

Any deviation from the SM expectations in measured production rates or any possible change in certain kinematic
distributions could  provide evidence for effects from physics beyond the SM. New physics processes at high mass scales
that alter the $\WW$ production can be described by
operators with mass dimensions larger than four in an effective field theory (EFT) framework. The higher-dimensional
operators of the lowest order from purely electroweak processes have dimension six, and
can generate anomalous trilinear gauge couplings (ATGC)~\cite{eftbasis}. Thus the measurement of the
coupling constants provides an indirect search for new physics at mass scales not directly accessible by the LHC.
Aside from the tests of the SM, $\WW$ production represents an important background source in searches for new
particles, and its precise measurement is therefore important in searches for new physics.

This paper is organized as follows. After a brief description of the CMS detector in Section~\ref{sec:detector} and
of the data and simulated samples in Section~\ref{sec:samples}, the event reconstruction and selection is detailed
in Section~\ref{sec:objects}. The background estimation is described in Section~\ref{sec:backgrounds}, followed
by an estimate of the uncertainties in Section~\ref{sec:systematics}. Finally the results for the inclusive
$\WW$ production cross section and those in a given fiducial phase space are presented in Section~\ref{sec:WWxsec}.
The normalized differential cross sections are shown in Section~\ref{sec:diffWWxsec} and limits on ATGCs in
Section~\ref{sec:aTGC}. A summary is given in Section~\ref{sec:conclusions}.
\par}

\section{The CMS detector}\label{sec:detector}
The CMS detector, described in detail in Ref.~\cite{CMSdetector}, is a
multipurpose apparatus designed to study high \pt
physics processes in proton-proton and heavy-ion collisions.
A superconducting solenoid occupies the central region of the CMS detector, providing a magnetic
field of 3.8\unit{T} parallel to the beam direction. Charged-particle
trajectories are measured by the silicon pixel and strip trackers, which
cover a pseudorapidity region of $\abs{\eta} < 2.5$.
The crystal electromagnetic calorimeter (ECAL), and
the brass/scintillator hadron calorimeter surround the tracking volume
and cover $\abs{\eta} < 3$. The steel/quartz-fiber Cherenkov hadron forward (HF)
calorimeter extends the coverage to $\abs{\eta} < 5$. The muon system consists of gas-ionization
detectors embedded in the steel flux-return yoke outside the solenoid, and
covers $\abs{\eta} < 2.4$. The first level of the CMS trigger system (level 1),
composed of custom hardware processors, is designed to select the most
interesting events in less than 4\rm{\mus}, using information from the
calorimeters and muon detectors. The level 1 output rate is up to 100\unit{kHz}.
The high-level trigger processor farm further reduces the event rate to
a few hundred Hz before data storage.

\section{Data and simulated samples}\label{sec:samples}
The data samples used correspond to an integrated luminosity
of 19.4\fbinv at $\sqrt{s} = 8\TeV$. The luminosity is measured using data from the HF system and the
pixel detector~\cite{lumiPAS2012}.

Events are selected with a combination of triggers that require
one or two high-\pt electrons or muons with relatively tight lepton
identification, some of them including also isolation. The single-electron trigger \pt
threshold is 27\GeV whereas that for single muons  is
24\GeV. For the dilepton triggers, the \pt thresholds of the leading and
trailing leptons are 17 and 8\GeV, respectively. The trigger efficiency is
measured in data using $\Z\to \ell^+\ell^-$ events recorded with a dedicated unbiased
trigger~\cite{wzxs}. The overall trigger efficiency is over 98\% for signal
events from $\cPq\cPaq\to\PWp\PWm$ and $\cPg\cPg \to \PWp\PWm$ processes within
our kinematic and selection region. The trigger efficiency is measured as a function
of the lepton \pt and $\eta$. In addition, prescaled single-lepton
triggers with \pt thresholds of 8 and 17\GeV are used for some of the data-driven background
estimations.

Several Monte Carlo (MC) event generators are used to simulate the signal and
background processes. The MC samples are used to optimize the event selection,
evaluate efficiencies and acceptances, and to estimate yields. For all MC samples,
the response of the CMS detector is simulated using a detailed description
of the detector based on the \GEANTfour package~\cite{Agostinelli:2002hh}. The simulated events
are corrected for the trigger efficiency to match the data.

The $\cPq\cPaq\to\PWp\PWm$ component of the signal is generated with
\POWHEG 2.0~\cite{powheg1,powheg2,powheg3,powheg4,powheg5}. For
comparison we also use $\cPq\cPaq\to\PWp\PWm$ signal samples generated with
the \MADGRAPH 5.1~\cite{madgraph} and \MCATNLO 4.0~\cite{MCatNLO} event
generators. The $\cPg\cPg \to \PWp\PWm$ signal component is generated using
\textsc{gg2ww} 3.1~\cite{ggww}. The sum of the $\cPq\cPaq\to\PWp\PWm$ and
$\cPg\cPg \to \PWp\PWm$ components is normalized to the inclusive $\ppww$
cross section at NNLO~\cite{NNLOXsec} accuracy.

Background processes with top quarks, $\ttbar$ and $\PQt\PW$, are generated with
\POWHEG. Higgs boson processes are considered part of the background. They represent
about 8\% of the $\WW$ cross section at $\sqrt{s} = 8\TeV$~\cite{NNLOXsec}, but have a
smaller signal efficiency and represent only about 3\% of the expected signal
yield. The gluon fusion and vector boson fusion modes are generated with
\POWHEG for a Higgs boson mass of 125\GeV and normalized to the SM
cross section~\cite{LHCHiggsCrossSectionWorkingGroup:2011ti}. The simulation of associated
Higgs production uses the
\PYTHIA 6.4 generator~\cite{pythia}. The interference between the Higgs boson production process and the $\WW$ continuum process is found to be approximately 0.1\%; the interference is significant only with
the $\cPg\cPg \to \PWp\PWm$ process. The $\PW\Z$, $\Z\Z$, $\V\V\V$
($\V = \PW/\Z$), \dyll, $\PW\gamma^{*}$, and \PW+jets processes are generated
using \MADGRAPH. All other background processes are generated using \PYTHIA 6.4.

The set of parton distribution functions (PDF) used is CTEQ6L~\cite{cteq66}
for leading order (LO) generators and CT10~\cite{Lai:2010vv} for
next-to-leading-order (NLO) generators. All the event generators are
interfaced to \PYTHIA 6.4 for the showering and hadronization of partons, except \MCATNLO, which is interfaced to
\HERWIG 6~\cite{herwig}. The \TAUOLA 2.7 package~\cite{tauola} is used in the simulation of $\tau$
decays to account for polarization effects.

\label{sec:samples:higherorder}
In order to suppress the top quark background processes, the $\ppww$ cross
section is measured with events that have no more than one high-\pt jet.
The veto on high-\pt jets enhances the importance of logarithms of the jet \pt, spoiling
the convergence of fixed-order calculations and requiring the use of dedicated
resummation techniques for an accurate prediction of differential distributions~\cite{ptresum,ptresum3}.
The \pt of the jets produced in association with the $\WW$ system is strongly correlated
with the transverse momentum of the $\WW$ system, $\pt^{\PW\PW}$, especially in the case where only one jet is produced. Thus,
a precise modeling of the $\pt^{\PW\PW}$ distribution
is necessary for the estimation of the jet veto efficiency. In
Ref.~\cite{ptresum}, the logarithmic terms that contribute to the $\pt^{\PW\PW}$
distribution from $\cPq\cPaq \to \PWp\PWm$ are resummed to
next-to-next-to-leading-logarithm precision using the technique of
\pt resummation~\cite{ptresum2}. The simulated $\cPq\cPaq \to \PWp\PWm$ signal events are reweighted according
to the ratio of the $\pt^{\PW\PW}$ distribution from the \pt-resummed calculation
and from \POWHEG and \PYTHIA. An equivalent reweighting procedure is applied to
\MCATNLO and \MADGRAPH MC generators. The weights have different effects for each MC generator; the
change in the jet veto efficiency estimated with \POWHEG is about 3\% whereas it is 1\% for
\MCATNLO and 4\% for \MADGRAPH. We find good agreement between the jet veto efficiency
estimated with \POWHEG, \MCATNLO, and \MADGRAPH after the equivalent
reweighting procedure is applied to these MC generators.

Additional simulated proton-proton interactions overlapping with the event of
interest, denoted as pileup events, are added to the simulated samples to reproduce
the vertex multiplicity distribution measured in data. The average value of pileup
events per bunch crossing is approximately 21.

\section{Event reconstruction and selection}\label{sec:objects}

A particle-flow algorithm~\cite{PFT-09-001,PFT-10-001} is used to reconstruct the
observable particles in the event by an optimized combination of  information
from different subdetectors: clusters of energy deposits measured
by the calorimeters and charged-particle tracks identified in the central tracking system
and the muon detectors.

This analysis uses leptonic decays $\PW \to \ell \nu$ ($\ell$ = $\Pe$, $\mu$), so the signal
candidates consist of three final states: $\Pep\Pem$, $\PGmp\PGmm$, and $\Pepm\PGmmp$. The signal
candidates contain a small contribution from $\PW\to\tau\nu_{\tau}$ processes with leptonic
$\tau$ decays, even though the analysis is not optimized for this final state.
The contribution of these leptonic $\tau$ decays to the final signal candidates is about 10\%.

For each signal event, two oppositely charged lepton candidates are required,
both with $\pt > 20\GeV$ and with $\abs{\eta} < 2.5 (2.4)$ for electrons (muons).
Among the vertices identified in the event, the vertex with the largest $\sum \pt^2$, where
the sum runs over all charged tracks associated with the vertex, is chosen as the primary one. The lepton
candidates are required to be compatible with originating from this primary vertex.

Electron candidates are defined by a reconstructed particle track in the
tracking detector pointing to a cluster of energy deposits in the ECAL.
A multivariate approach to identify electrons is employed~\cite{tmva} combining
several measured quantities describing the track quality, the ECAL cluster shape,
and the compatibility of the measurements from the two subdetectors.
The electron energy is measured primarily from the ECAL cluster energy deposit~\cite{ElectronReco}.
Muon candidates are identified by signals of particle tracks in the muon system
that match a track reconstructed in the central tracking system.
Minimum requirements on the number of hits and on the goodness-of-fit of the full track
are imposed on the muon curvature measurement~\cite{MuonReco}.

The signal electrons and muons are required to be isolated to distinguish them from the semileptonic decays of heavy quarks or the in-flight decays of hadrons.
The $\Delta R = \sqrt{\smash[b]{(\Delta\eta)^2 + (\Delta\phi)^2}}$ variable is used to measure
the separation between reconstructed objects in the detector, where $\phi$
is the azimuthal angle (in radians) of the trajectory of the object in
the plane transverse to the direction of the proton beams, and therefore
$\Delta\phi$ is the $\phi$ separation between objects; $\Delta\eta$ is the
$\eta$ separation between objects.
Isolation criteria are set based on the distribution of low-momentum particles
in the ($\eta,\phi$) region around the leptons. To remove the contribution from the
overlapping pileup interactions in this isolation region,
the charged particles included in the computation of the isolation variable
are required to originate from the primary vertex. This track assignment to the primary vertex
is fairly loose, and includes most of the tracks from b-quark or c-quark decays. The neutral
component in the isolation $\Delta R$ cone is corrected by the average energy density
deposited by those neutral particles that originated from additional interactions~\cite{Cacciari:subtraction}. The correction is measured in a region of the detector away from the known hard scattering in a control sample.

Electron isolation is characterized by the ratio of
the total \pt of the particles reconstructed in a $\Delta R =0.3$ cone
around the electron, excluding the electron itself,
to the \pt of the electron.
Isolated electrons are selected by requiring this ratio to be
below 10\%. For each muon candidate, the scalar sum of the \pt of all particles
originating from the primary vertex is reconstructed in
$\Delta R$ cones of several radii around the muon direction, excluding the
contribution from the muon itself. This information is combined using a
multivariate algorithm that exploits the differential energy deposition in the
isolation region to discriminate between the signal of prompt muons and muons
from hadron decays inside a jet. The exact threshold value depends on the muon
$\eta$ and \pt~\cite{PublishedHWW}.

Jets are reconstructed using the anti-\kt clustering algorithm~\cite{antikt}
with a distance parameter of 0.5, as implemented in the \FASTJET
package~\cite{Cacciari:fastjet1,Cacciari:fastjet2}. The properties of the jets
are modified by particles from pileup interactions. A combinatorial background arises from
low-\pt jets from pileup interactions, which are clustered together with high-\pt jets
from the primary interaction. A multivariate jet identifier is applied to separate jets from the
primary interaction and those reconstructed from energy deposits associated with
pileup interactions~\cite{jetIdPAS}. The discrimination is based on the differences in the jet shapes,
on the relative multiplicity of charged and neutral components,
and on the different
\pt fractions carried by the hardest components.
Tracks that come from pileup vertices are removed from the jet clustering.
After jet identification, we apply a correction similar to
the one applied for lepton isolation that accounts for the contributions from
pileup.
Jet energy corrections are applied as a function
of the jet \pt and $\eta$~\cite{cmsJEC}.
Studies of the jet multiplicity as a function of the
number of vertices have been performed using Z+jets events,
and no significant dependence was found.
Since the jet energy resolution in data is somewhat worse than in simulation, the \pt values of simulated jets
need to be spread randomly 5\% in order to describe data.
After corrections the jets considered for the event categorization are required to have $\pt>30\GeV$ and $\abs{\eta}<4.7$.

To reduce the background from top quark decays, events with two or more jets
surviving the jet selection criteria are rejected.
To further suppress the top quark background, two tagging techniques based
on soft-muon and b-quark jet tagging are applied~\cite{btag}. The first method vetoes
events containing a soft muon from the semileptonic decay of the b quark. Soft-muon
candidates are defined without isolation requirements and are required to have $\pt > 3\GeV$.
The second method uses b-jet tagging criteria based on the impact parameter
of the constituent tracks. In particular, a track counting high-efficiency  algorithm is used
to veto those events with a jet tagged as b quark (t-tagged events).
The combined reduction of the top quark background is about 50\% in the zero-jet category
and above 80\% for events with one jet with $\pt >30\GeV$.

The $\VEtmiss$ variable is defined as the negative vector sum of
the \pt of all reconstructed particles (charged or neutral) in the event.
A \textit{projected}~\MET variable~\cite{PublishedHWW} is defined as the component of $\VEtmiss$ transverse
to the nearest lepton if the lepton is situated within an azimuthal angular window of ${\pm} \pi/2$ from
the $\VEtmiss$ direction, otherwise the $\abs{\VEtmiss}$ is used.
This variable is particularly effective in rejecting
(1) $\dytt$ events where $\VEtmiss$ is preferentially aligned with leptons, and (2)  $\dyll$
events with poorly measured $\VEtmiss$. Since
the $\VEtmiss$ resolution is degraded in a high pileup environment, two projected~\MET variables are defined: one
constructed from all identified particles (proj. \MET), and another constructed
from the charged particles attached to the primary vertex only (proj. track \MET). The minimum of the two is required
to be above 20\GeV.

Events with dilepton masses below 12\GeV are also rejected
to remove contributions from low-mass resonances. The same
requirement is applied to the $\Pepm\PGmmp$ final state to reject multijet and $\PW\gamma$ background processes. Finally, the transverse
momentum of the dilepton system $\pt^{\ell\ell}$ is required to be above 45\GeV
in the $\Pep\Pem$ and $\PGmp\PGmm$ final states, and above 30\GeV in the $\Pepm\PGmmp$
final state to reduce both the Drell--Yan background and events containing jets misidentified as leptons.

The Drell--Yan (DY) $\Z/\gamma^*$ process is the largest source of
same-flavor lepton pair production background because of its large production
cross section and the finite resolution of the $\VEtmiss$ measurement.
In order to suppress this background, a few additional selection
requirements are applied to the same-flavor final states.
The component of the Drell--Yan production close to the Z boson peak is rejected by requiring the dilepton invariant mass $\mll$ to be more than 15\GeV away
from the $\Z$ boson mass. To suppress the remaining off-peak contribution,
a dedicated multivariate selection is used, combining \MET variables, kinematic variables of the dilepton system, the transverse  mass, the leading jet \pt, and differences in azimuthal angle between the dilepton system and the leading jet and the $\VEtmiss$~\cite{PublishedHWW}.
These selection requirements effectively reduce the Drell--Yan background by three orders of magnitude,
while retaining more than 50\% of the signal.

To reduce the background from other diboson processes, such as $\WZ$ and $\ZZ$
production, any event that has an additional third lepton
passing the identification and isolation requirements and having  $\pt > 10 \GeV$
is rejected.
Any $\PW\gamma$ production where the photon converts is suppressed
by rejecting electrons consistent with a photon conversion~\cite{ElectronReco}.

A summary of the selection requirements for different- and same-flavor final states
is shown in Table~\ref{tab:event_sel}.

\begin{table*}[htbp]
  \begin{center}
 \topcaption{Summary of the event selection for the different-flavor and
  same-flavor final states.\label{tab:event_sel}}
  \begin{tabular} {lcc}
  \hline
 Variable & Different-flavor & Same-flavor \\
  \hline
Opposite-sign charge requirement            & Applied & Applied \\
$\pt^{\ell}$ [\GeVns{}]                       & $>$20 & $>$20 \\
min(proj. \MET, proj. track \MET)[\GeVns{}]             & $>$20 & $>$20    \\
\multirow{2}{*}{DY MVA}                     & \multirow{2}{*}{\NA}        &  $>$0.88 in zero-jet  \\
                                            &                           & ($>$0.84 in one-jet) \\
$\abs{\mll - m_{\Z}}$ [\GeVns{}]                  & \NA           & $>$15 \\
$\pt^{\ell\ell}$ [\GeVns{}]                   & $>$30 & $>$45 \\
$\mll$ [\GeVns{}]                             & $>$12 & $>$12 \\
Additional leptons ($\pt^{\ell} > 10\GeV$) & veto        & veto \\
Top-quark veto                              & applied     & applied  \\
Number of reconstructed jets                & $<$2         & $<$2    \\
\hline
  \end{tabular}
  \end{center}
\end{table*}

\section{Estimation of backgrounds}\label{sec:backgrounds}

\begin{figure*}[htpb]
\centering
\includegraphics[width=0.45\textwidth]{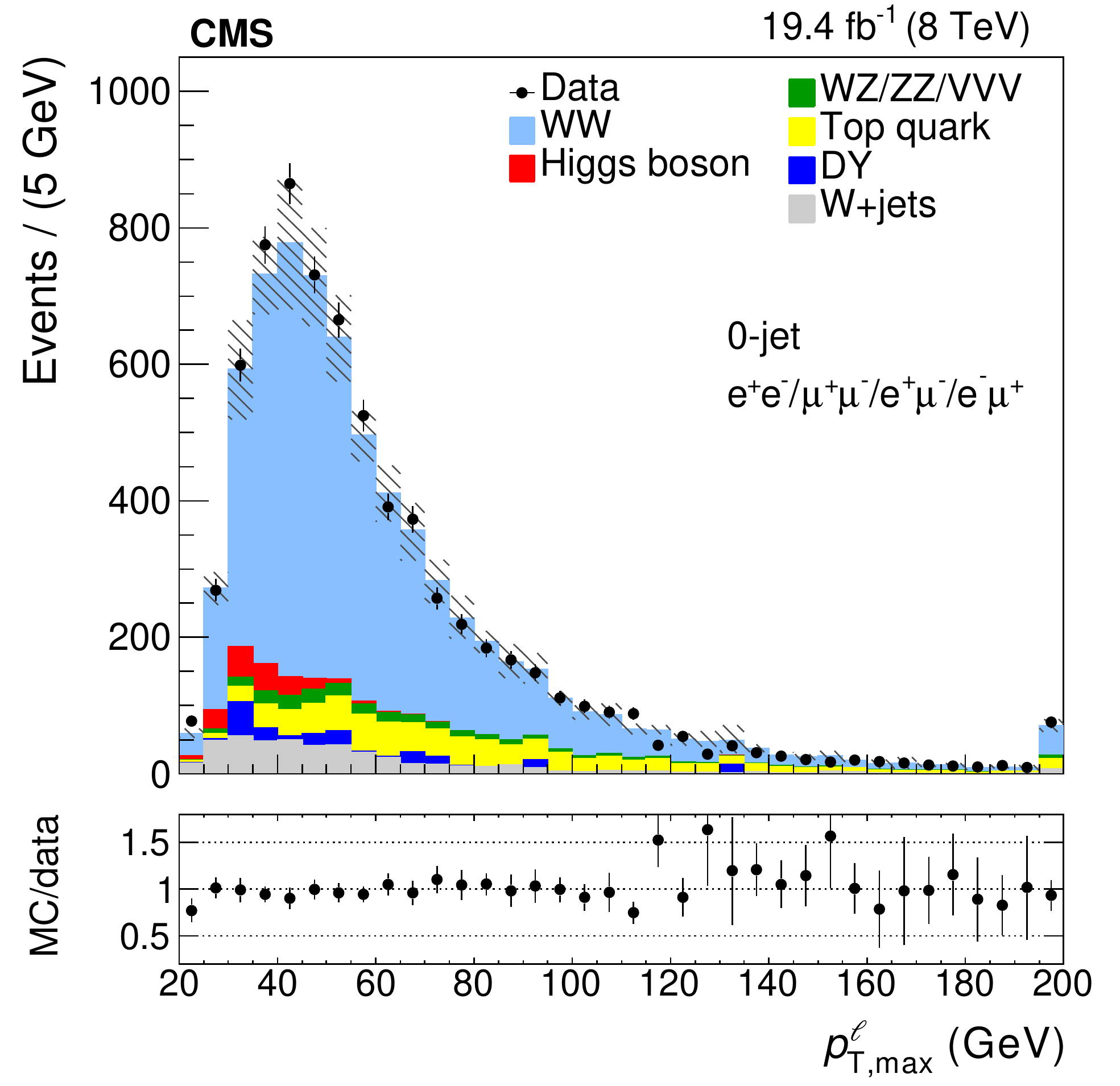}
\includegraphics[width=0.45\textwidth]{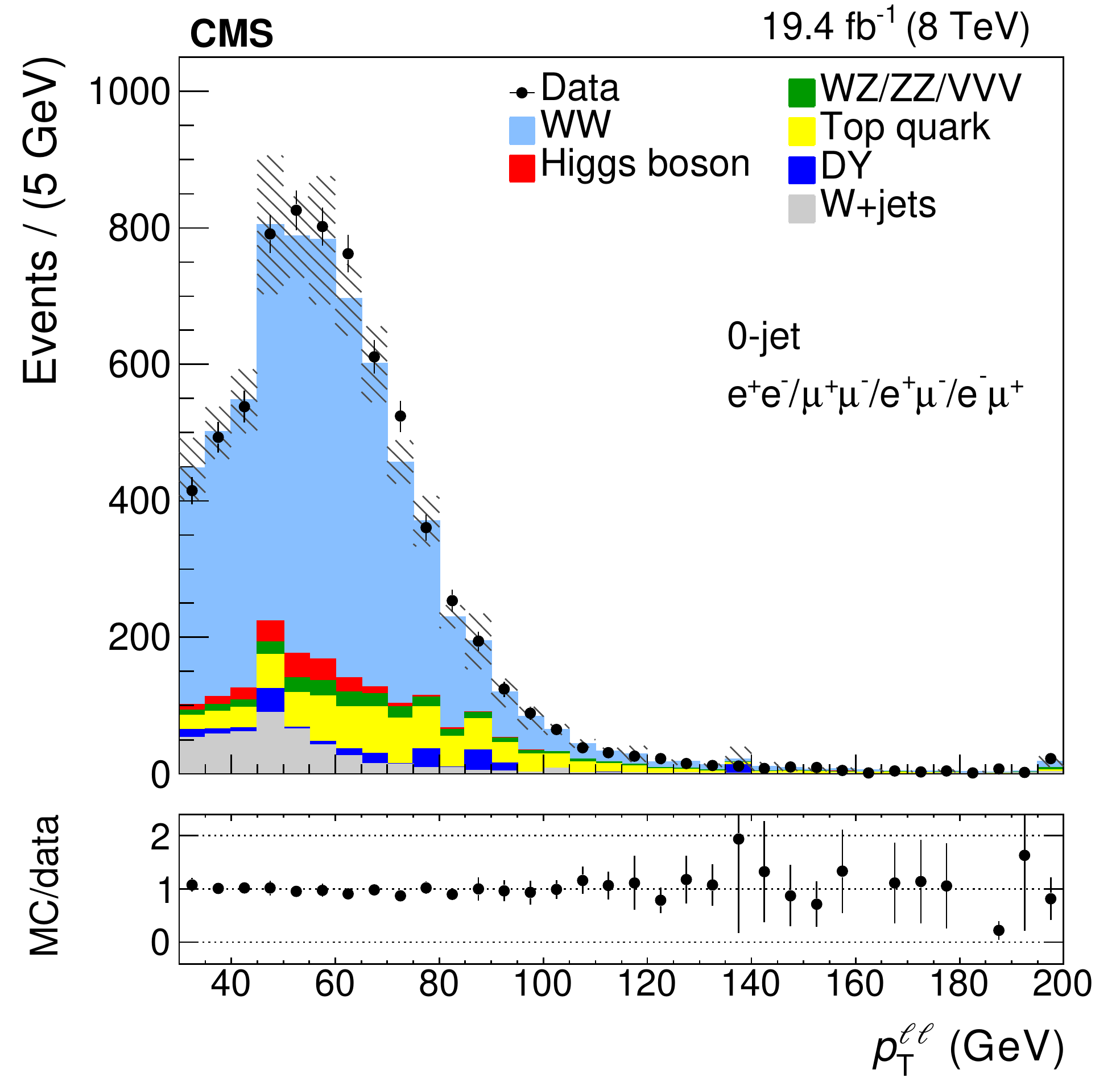}\\

\includegraphics[width=0.45\textwidth]{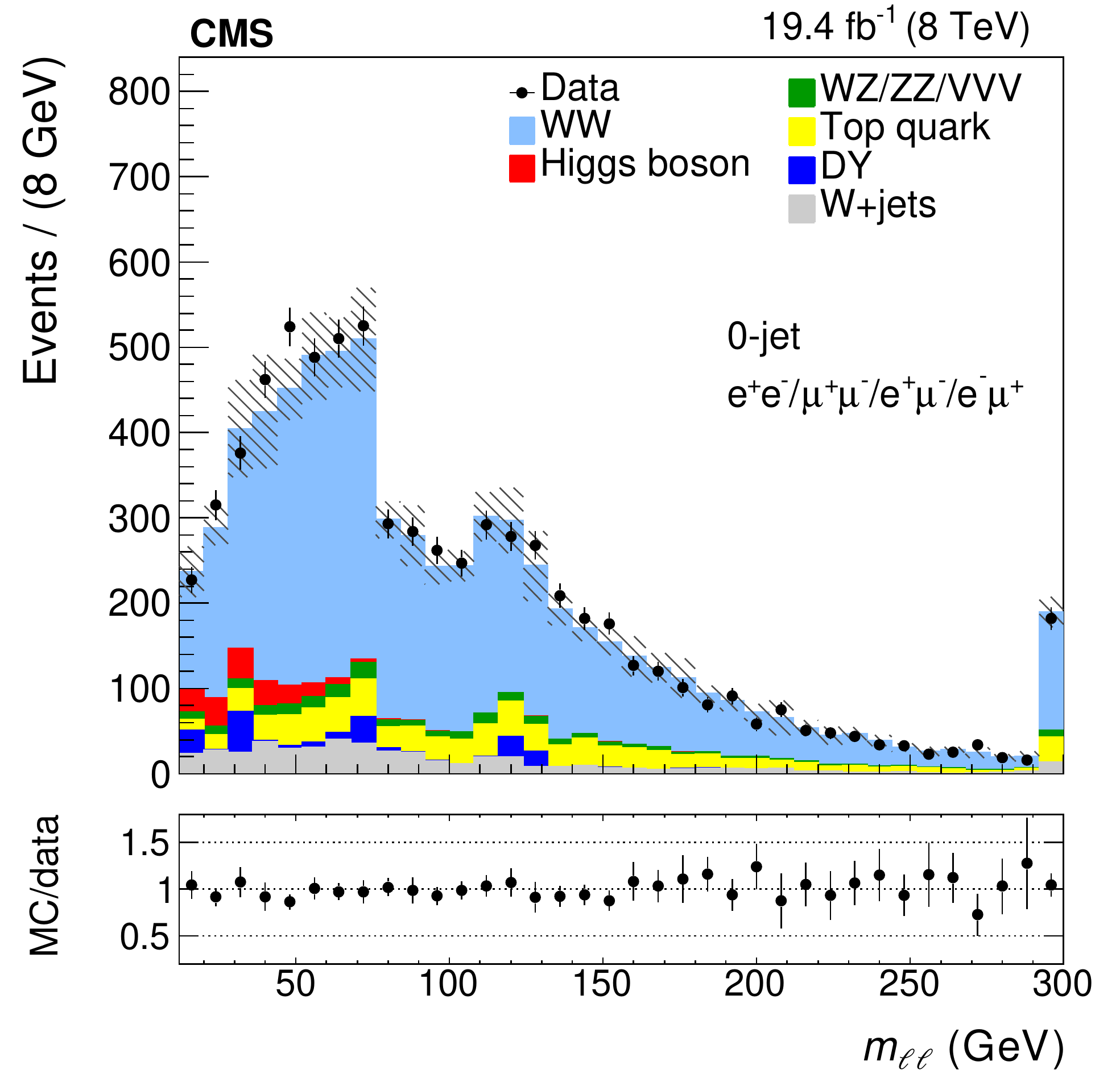}
\includegraphics[width=0.45\textwidth]{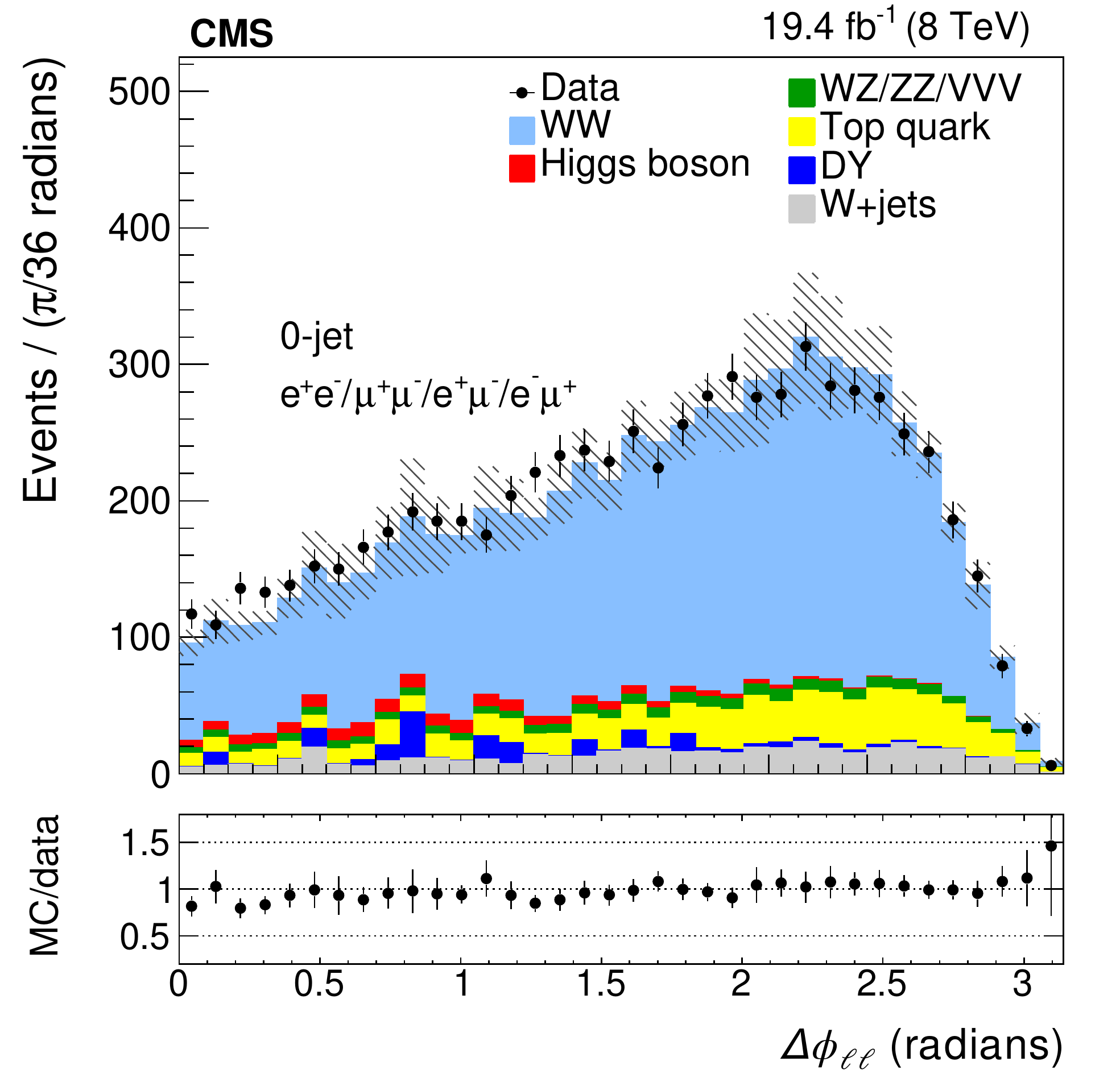}
\caption{\label{fig:WWxsec:1} The data and MC distributions for the zero-jet category of the leading lepton \pt ($p_{{\mathrm{T}},\text{ max}}^{\ell}$), the \pt of the dilepton system ($\pt^{\ell\ell}$), the dilepton invariant mass ($m_{\ell\ell}$) and the azimuthal angle between the two leptons ($\Delta\phi_{\ell\ell}$). The hatched areas represent the total systematic uncertainty in each bin. The error bars in the ratio plots are calculated considering the statistical uncertainty from the data sample and the systematic uncertainties in the background estimation and signal efficiencies. The last bin includes the overflow.}
\end{figure*}

\begin{figure*}[htpb]
\centering
\includegraphics[width=0.45\textwidth]{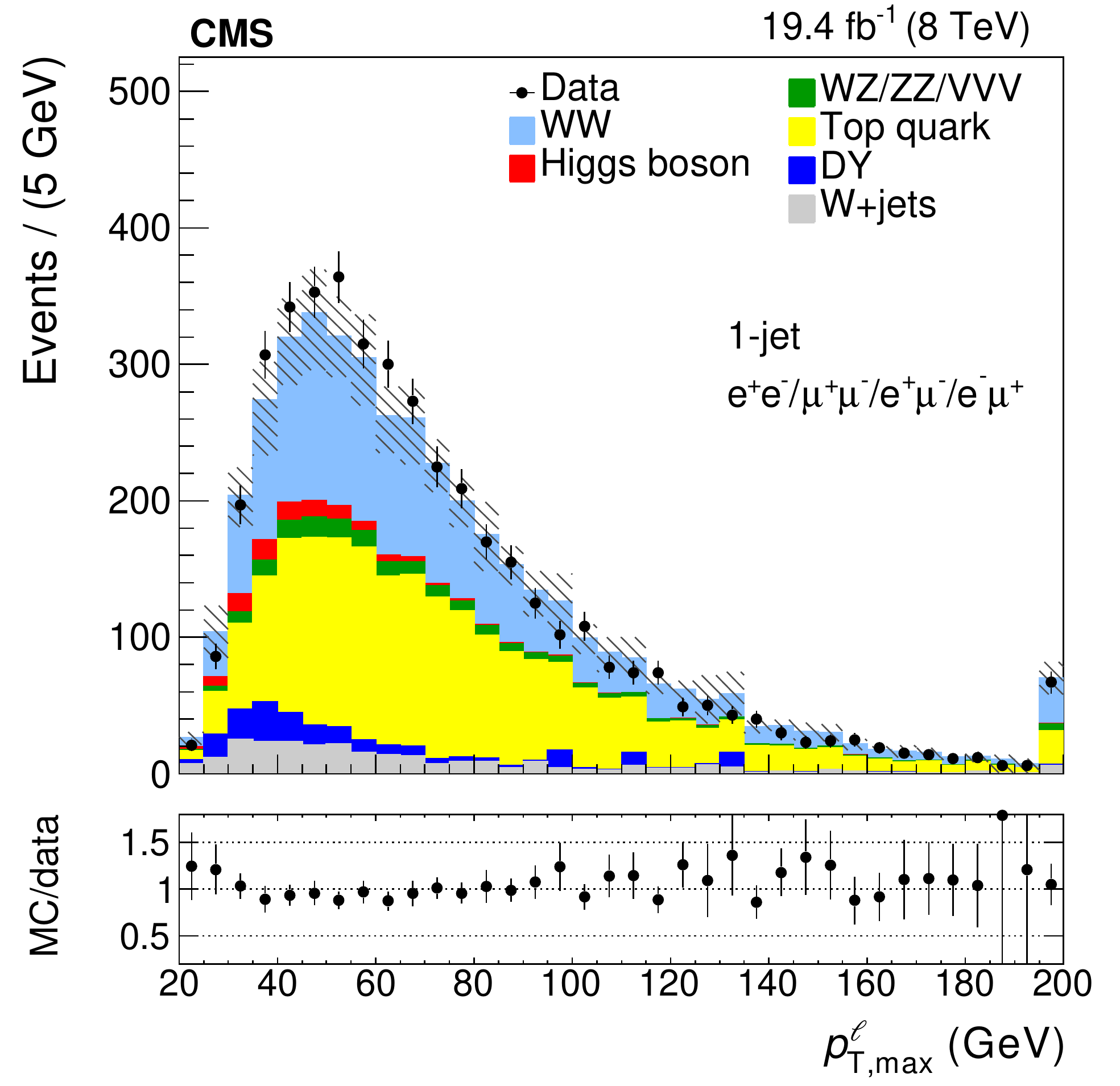}
\includegraphics[width=0.45\textwidth]{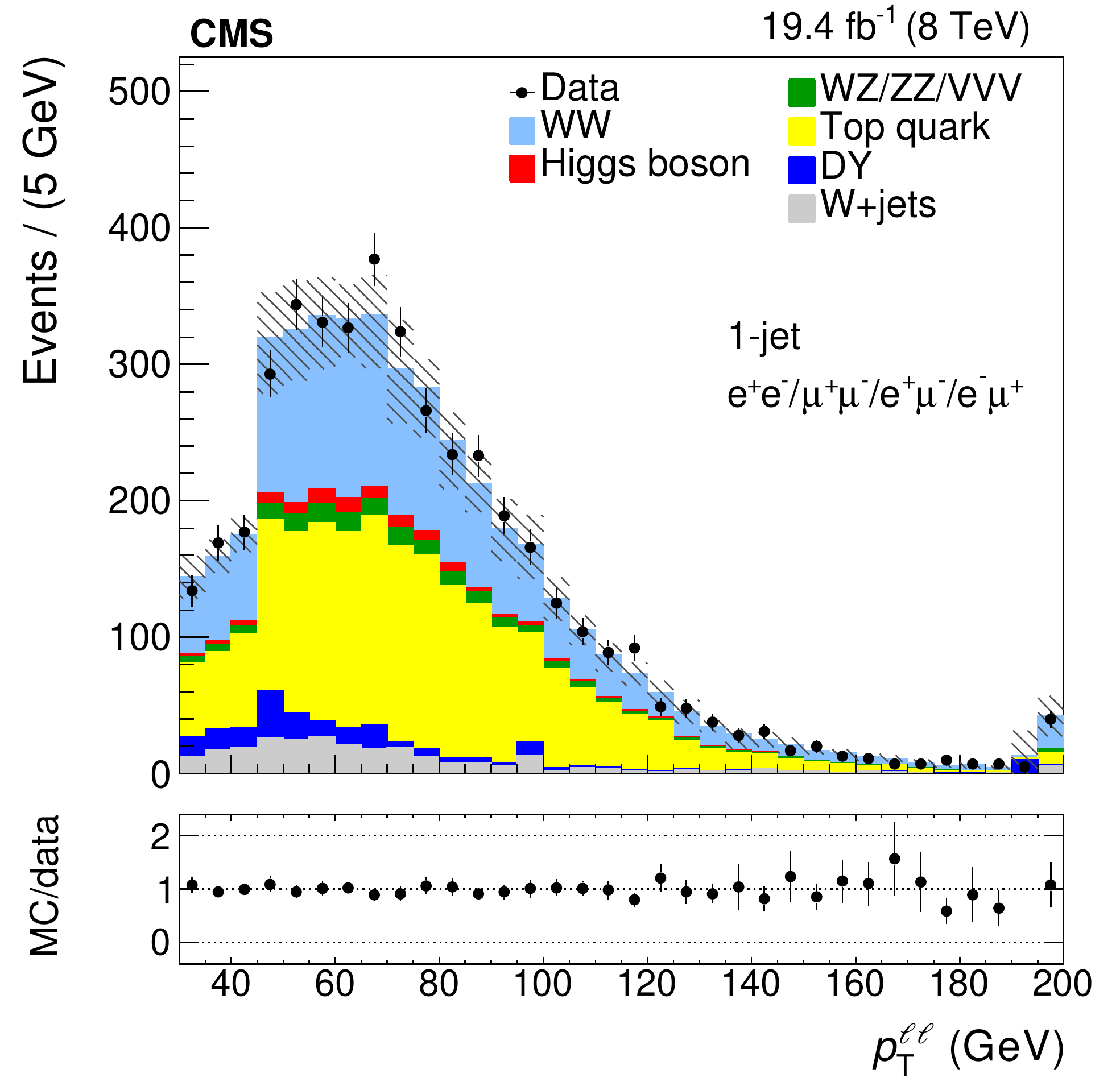}\\

\includegraphics[width=0.45\textwidth]{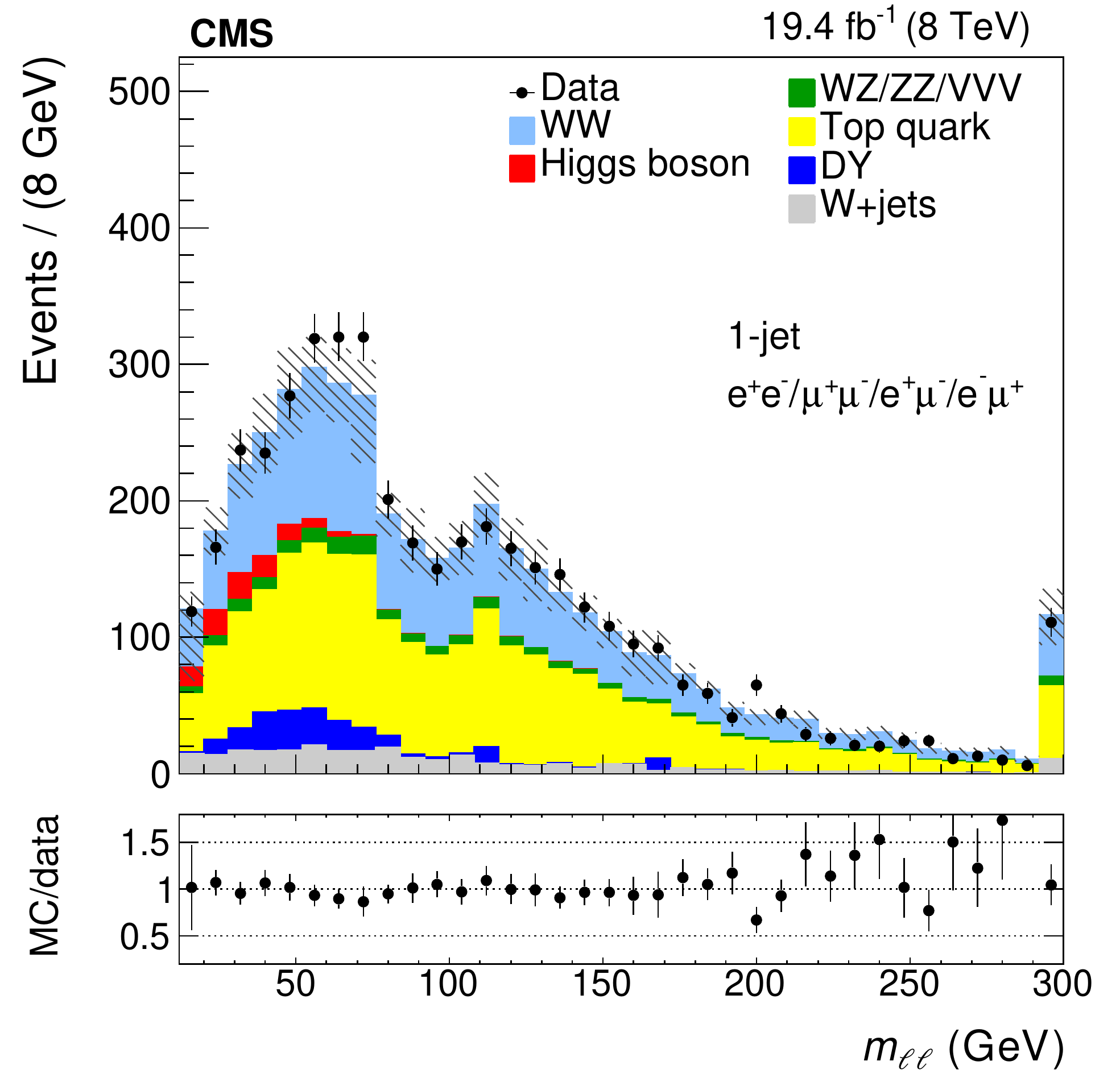}
\includegraphics[width=0.45\textwidth]{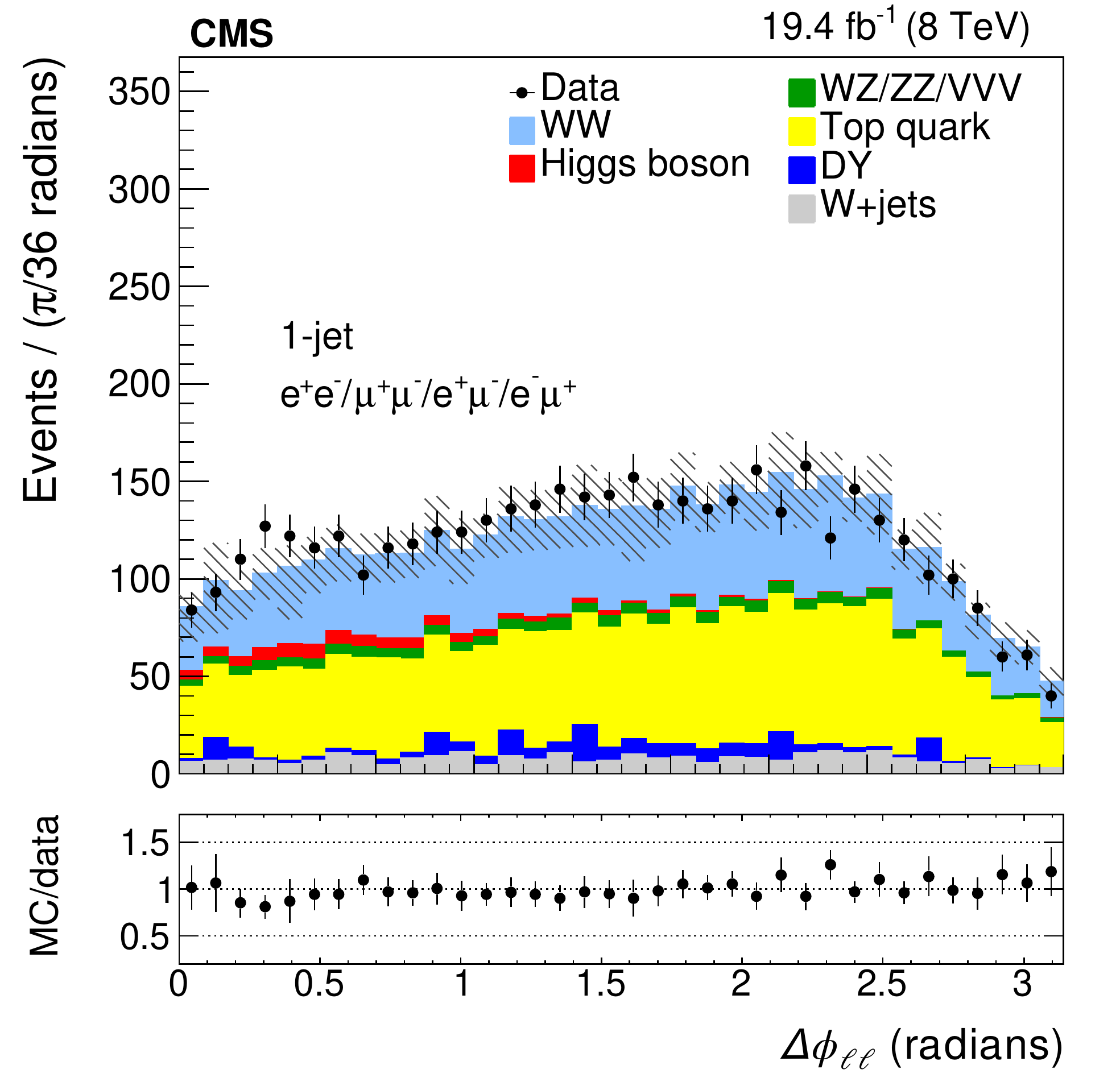}
\caption{\label{fig:WWxsec:2} The data and MC distributions for the one-jet category of the leading lepton \pt ($p_{{\rm{T}},\text{ max}}^{\ell}$), the \pt of the dilepton system ($\pt^{\ell\ell}$), the dilepton invariant mass ($m_{\ell\ell}$) and the azimuthal angle between the two leptons ($\Delta\phi_{\ell\ell}$). The hatched areas represent the total systematic uncertainty in each bin. The error bars in the ratio plots are calculated considering the statistical uncertainty from the data sample and the systematic uncertainties in the background estimation and signal efficiency. The last bin includes the overflow.}
\end{figure*}

A summary of the data, signal, and background yields for the different event categories is
shown in Table~\ref{tbl:WWxsec:1}. The distributions of the leading lepton \pt ($p_{{\mathrm{T}},\text{ max}}^{\ell}$), the \pt of the dilepton system ($\pt^{\ell\ell}$), the dilepton invariant mass ($m_{\ell\ell}$) and the azimuthal angle between the two leptons ($\Delta\phi_{\ell\ell}$) are shown in Figs.~\ref{fig:WWxsec:1} and ~\ref{fig:WWxsec:2} for the zero-jet and one-jet categories.

\begin{table*}[htbp]
\centering
\topcaption{\label{tbl:WWxsec:1}Data, signal, and background yields for the four
different event categories used for the $\ppww$ cross section measurement.
The reported uncertainties include both statistical and systematic components as described in Section~\ref{sec:systematics}.}
{\small
\begin{tabular}{lxxxx}
\hline
\multirow{2}{*}{Process}                 &  \multicolumn{2}{c}{zero-jet category} & \multicolumn{2}{c}{one-jet category} \\
                                         & \multicolumn{1}{c}{Different-flavor}      & \multicolumn{1}{c}{Same-flavor}   &  \multicolumn{1}{c}{Different-flavor} &  \multicolumn{1}{c}{Same-flavor}             \\
\hline
$\cPq\cPaq\to\PWp\PWm$  &  3516 , 271   &   1390 , 109   &     1113 , 137   &  386 ,  49 \\
$\Pg\Pg\to\PWp\PWm$     &   162 ,  50   &     91 ,  28   &       62 ,  19   &   27 ,   9 \\
\hline
$\PWp\PWm$              &  3678 , 276   &   1481 , 113   &     1174 , 139   &  413 ,  50 \\
\hline
$\ZZ+\WZ$               &    84 ,  10   &     89 ,  11   &       86 ,   4   &   42 ,   2 \\
$\V\V\V$                &    33 ,  17   &     17 ,   9   &       28 ,  14   &   14 ,   7 \\
top quark (${\mathrm{B}}_\text{t-tag}$)                 &   522 ,  83   &    248 ,  26   &     1398 , 156   &  562 , 128 \\
$\dyll$                 &    38 ,   4   &    141 ,  63   &      136 ,  14   &   65 ,  33 \\
$\PW\gamma^{*}$          &    54 ,  22   &     12 ,   5   &       18 ,   8   &    3 ,   2 \\
$\PW\gamma$              &    54 ,  20   &     20 ,   8   &       36 ,  14   &    9 ,   6 \\
$\PW+\text{jets}(\Pe)$             &   189 ,  68   &     46 ,  17   &      114 ,  41   &   16 ,   6 \\
$\PW+{\text{jets}}{(\mu)}$           &    81 ,  40   &     19 ,   9   &       63 ,  30   &   17 ,   8 \\
Higgs boson                   &   125 ,  25   &     53 ,  11   &       75 ,  22   &   22 ,   7 \\
\hline
Total bkg.              &  1179 , 123   &    643 ,  73   &     1954 , 168   &  749 , 133 \\
\hline
$\PWp\PWm$ + total bkg. &  4857 , 302   &   2124 , 134   &     3128 , 217   & 1162 , 142 \\
\hline
Data                    &  \multicolumn{1}{c}{4847}  &   \multicolumn{1}{c}{2233}  &   \multicolumn{1}{c}{3114} &  \multicolumn{1}{c}{1198}           \\
\hline
\end{tabular}
}
\end{table*}

A combination of techniques is used to determine the contributions from
backgrounds that remain after the $\WW$ selection. A detailed description
of these techniques can be found in Ref.~\cite{PublishedHWW}.
The main background comes from top quark production, which is
estimated from data. Instrumental backgrounds arising from
misi\-denti\-fi\-ed (``nonprompt") leptons in \PW+jets production and
mismeasurement of $\VEtmiss$ in $\Z/\gamma^*$+jets events are also estimated from
data. Other contributions from $\PW\gamma$, $\PW\gamma^*$, and other subdominant diboson
($\PW\Z$ and $\Z\Z$) and triboson ($\V\V\V$) production processes
are estimated partly from simulated samples.

A common scale factor is estimated for the $\ttbar$ and tW simulated samples. The top-quark background is suppressed using a top-tagging veto that eliminates visible top-quark decays. After the full event selection described in Table~\ref{tab:event_sel} but before the top-quark veto, the remaining top-quark background contribution ($\mathrm{B}_\text{t-tag}$) is estimated as: ${\mathrm{B}}_\text{t-tag} = {N}_\text{t-tag} \, (1-\epsilon_\text{t-tag})/\epsilon_\text{t-tag}$, where ${N}_\text{t-tag}$ is the number of t-tagged events before the top-quark veto, and $\epsilon_\text{t-tag}$ is the corresponding t-tagged efficiency. The number of t-tagged events (${N}_\text{t-tag}$) is determined in the signal data sample by counting the number of events passing the t-tagging requirements described in Section~\ref{sec:objects} and subtracting any remaining background on the basis of simulations or data, as described in the present section. The t-tagged efficiency ($\epsilon_\text{t-tag}$) is obtained from a measurement of the efficiency to tag a b-quark jet or soft muon in a top-enriched sample that consists of events with one (two) jet and exactly one b-tagged jet with $\pt>30$\GeV, which isolates one b quark in a sample that is primarily $\ttbar$ or $\PQt\PW$ events. Any remaining background is subtracted from the measured data in the top-enriched control sample. After excluding this b-tagged jet, the t-tagging efficiency is determined by counting the number of events that have an additional b-tagged jet or a soft muon. The measured efficiency is defined per b-quark decay and the value measured in the top-enriched sample is converted to a top-tagging efficiency in the signal region by taking into account the relative difference in the number of b-quark jets between the two samples after excluding the high-$\pt$ b-tagged jet used to select events in the control sample. The conversion factor is calculated using the ratio of expected single-top $\PQt\PW$ events to top-quark pair $\ttbar$ events in each region, and is done separately for the 0-jet and 1-jet categories as described in detail in Appendix D of Ref.~\cite{PublishedHWW}. We obtain efficiency values of 50--70\% in the signal samples. The main uncertainty comes from the statistical uncertainty in the control sample and from the systematic uncertainties related to the measurement of $\epsilon_\text{t-tag}$. The total uncertainty in ${\rm{B}}_\text{t-tag}$ amounts to about 13\% in the zero-jet category and 3\% in the one-jet category. The top background estimation method gives the estimate for the count of events in each of the 4 channels. This estimate is used to normalize the integral of the simulated distributions of $\ttbar$ and $\PQt\PW$ backgrounds used in this paper.

The nonprompt lepton background occurs in  \PW+jets and dijets production and originates from leptonic decays of heavy quarks, hadrons
misi\-denti\-fi\-ed as leptons, and electrons from photon conversion. Most of it is suppressed by the identification and isolation requirements
on electrons and muons described in Section~\ref{sec:objects}. The remaining contribution
is estimated directly from data from a sample enriched in nonprompt leptons. This sample is
selected by choosing events with one lepton candidate that passes the standard lepton selection criteria, and another
lepton candidate that fails the criteria, but passes a looser selection on impact parameter and isolation resulting in a sample of ``pass-fail" lepton pairs. The yield in this sample is extrapolated to the signal region using the efficiencies for such loosely identified leptons to pass the standard lepton selection criteria.

The efficiency, $\epsilon_\text{pass}$, for a jet that satisfies the loose lepton requirements to pass
the standard lepton selection is determined using an independent dijet sample. This independent dijet
sample consists of events with one lepton candidate passing loose selection criteria and a recoiling jet,
where contributions from $\PW$+jets and $\cPZ$+jets events are suppressed by rejecting events with
significant \MET or with additional leptons. In order to study the composition of the nonprompt background, different dijet samples are defined by requiring different jet-\pt thresholds for the jet recoiling against the  misidentified lepton. To ensure the measured efficiency is applicable to the signal region we compare
the \pt spectrum of the jets in the dijet sample, and in the pass-fail sample from which the extrapolation
is performed. The efficiency, parametrized as a function of \pt and $\eta$ of the lepton, is
used to weight the events in the pass-fail sample by $\epsilon_\text{pass}/(1 - \epsilon_\text{pass})$
to obtain the estimated contribution from the nonprompt lepton background in the signal region.
The systematic uncertainties from the determination of $\epsilon_\text{pass}$
dominate the overall uncertainty of this method. The systematic uncertainty is estimated
by modifying the jet \pt threshold in the dijets sample, which modifies the jet sample
composition, and from a closure test, where $\epsilon_\text{pass}$ is derived from simulated
dijet events and applied to simulated background samples to predict the number of background events.
The total uncertainty in $\epsilon_\text{pass}$ is of the order of 40\%, which includes
the statistical uncertainty arising from the control sample size.

The $\Z/\gamma^*\to \Pe\Pe/\mu\mu$ contribution, including $\Z/\gamma^*\to \tau\tau$
leptonic decays, in the same-flavor final states outside of the Z boson mass window is obtained
by normalizing the simulation. The normalization factor is defined by the ratio of the simulated to
the observed number of events inside the $\Z$ boson mass window in data. The contribution of
$\WZ$ and $\ZZ$  inside the $\Z$ boson mass window in data with neither lepton arising from a Z boson
is subtracted before performing the normalization. This is done by counting the number
of $\Pepm\PGmmp$ events in the Z mass window, accounting for combinatorial effects and
the relative detection efficiencies for electrons and muons. The contribution of $\WZ$
and $\ZZ$ processes in the Z mass window with leptons arising from different bosons, is also subtracted
as estimated from simulation. The largest uncertainty in the estimate arises from the dependence
of the extrapolation factor on \MET and the multivariate Drell--Yan discriminant.
The total uncertainty in the $\dyll$ normalization is about 30\%, including both statistical
and systematic components. The contribution of this background is also evaluated with an
alternative method using $\gamma$ + jets events, which provides results consistent with the
primary method. The $\dytt$ background in the $\Pepm\PGmmp$ channel is obtained from $\dymm$ events
selected in data, where the muons are replaced with simulated $\tau$ decays. The Drell--Yan event
yield is rescaled to the observed yield using the inclusive sample of $\dyll$~\cite{Htautau}.

A data sample with three reconstructed leptons is selected in order to normalize
the simulation used to estimate the $\PW\gamma^{*}$ background contribution coming
from asymmetric $\gamma^{*}$ decays where one lepton escapes detection~\cite{wgammastart}.
The systematic uncertainty is estimated by comparing the normalization factor estimated in
simulation  in different regions. The uncertainty in the $\PW\gamma^{*}$ background estimate
is of the order of 40\%.

Other backgrounds are estimated from simulation. The $\PW\gamma$ background simulation
is validated in data using the events passing all the selection requirements,
except that the two leptons must have the same charge; this sample is dominated by
\PW+jets and $\PW\gamma$ events. Differences in the overall normalization are counted
as a systematic uncertainty. The uncertainty in the $\PW\gamma$ background estimate is about 30\%.
Other minor backgrounds are $\WZ$ and $\ZZ$ diboson production where
the two selected leptons come from different bosons.

\section{Signal efficiency and systematic uncertainties}\label{sec:systematics}
The signal efficiency, which includes both detector geometrical acceptance and
signal reconstruction and selection efficiency, is estimated using the
$\cPq\cPaq\to\PWp\PWm$ and nonresonant (not through a Higgs resonance)
$\Pg\Pg\to\PWp\PWm$ signal simulations described in Section~\ref{sec:samples}.
Signal events from $\PW\to\tau\nu_{\tau}$ decays with $\tau$ leptons decaying
into lower-energy electrons or muons are included in the signal efficiency.
Residual discrepancies in the lepton reconstruction and identification
efficiencies between data and simulation are corrected by applying
data-to-simulation scale factors measured using $\dyll$ events in the $\cPZ$
peak region~\cite{wzxs} that are recorded with unbiased triggers. These factors
depend on the lepton \pt and $\eta$ and are within 2\% (4\%) for electrons (muons).
The uncertainty in the determination of the trigger efficiency leads to an uncertainty
of about 1\% in the expected signal yield. Any residual differences between the analysis
lepton requirements with respect to the trigger selections are covered by the
uncertainty in the trigger efficiency.

The experimental uncertainties in the lepton reconstruction and identification
efficiency, momentum scale and resolution, $\MET$ modeling, and jet energy scale
are applied to the reconstructed objects in simulated events by randomly spreading
and scaling the relevant observables and propagating the effects to the kinematic
variables used in the analysis. The distributions with varied detector response
and resolution are used to estimate the change in the signal efficiency, whose
value is taken as the associated systematic uncertainty. Uncertainties in lepton
momentum scale and resolution are 0.5--4\% per lepton depending on the kinematics,
and the effect on the yields at the analysis selection level is approximately 1\%.
The uncertainties in the jet energy scale and resolution result in a 2--3\%
uncertainty in the yields. The uncertainty in the resolution of the \MET
measurement is approximately 10\%, which is estimated from $\dyll$  events with
the same lepton selection as in the analysis. Randomly smearing the measured \MET
by one standard deviation of the resolution gives rise to 2\% variation in the
estimation of signal yields after the full selection. A 2.6\% uncertainty is
assigned to the integrated luminosity  measurement~\cite{lumiPAS2012}.

The relative uncertainty in the signal acceptance from variations of the PDFs
and the value of $\alpha_{s}$ in the simulated samples is estimated to be 1.3\%
(0.8\%) for $\cPq\cPaq$ ($\Pg\Pg$) production, following the {PDF4LHC}
prescription~\cite{pdf4lhcInterim,pdf4lhcReport,Lai:2010vv,MSTW2008pdf,NNPDFpdf,LHCHiggsCrossSectionWorkingGroup:2011ti}.
The effect of higher-order corrections in the $\cPq\cPaq\to\PWp\PWm$ signal
acceptance is studied using the $\pt^{\PW\PW}$ reweighting procedure described in
Section~\ref{sec:samples}. Uncertainties are estimated by performing the reweighting
while varying the resummation scale between half and twice the nominal value used
in Ref.~\cite{ptresum}. The reweighting functions with varied scales are then
applied to simulated \POWHEG events and used to calculate the variation in
the signal acceptance. Uncertainties in the $\cPq\cPaq\to\PWp\PWm$ signal acceptance
sensitive to the renormalization ($\mu_R$) and factorization ($\mu_F$) scales are
estimated by varying both scales in the range ($\mu_0/2,2\mu_0$), with $\mu_0$
equal to the mass of the $\PW$ boson, and setting $\mu_R = \mu_F$. The resummation
scale uncertainty is found to be 2.8\% (6.9\%) for the zero-jet (one-jet) selection.
The renormalization and factorization scales uncertainty is found to be 2.5\% (6.3\%)
for the zero-jet (one-jet) selection. The systematic uncertainty associated with
higher-order corrections to the $\Pg\Pg\to\PWp\PWm$ component of the signal is
estimated by varying the renormalization and factorization scales and is found
to be about 30\%.

The systematic uncertainties due to the underlying event and parton shower model
are estimated by comparing samples with different MC event generators. In particular,
the \POWHEG MC generator interfaced with \PYTHIA for the parton shower and
hadronization is compared to the \MCATNLO generator interfaced with \HERWIG
for the parton shower and hadronization model. The systematic uncertainty is
found to be 3.5\%.

The uncertainties in the background predictions are described in Section~\ref{sec:backgrounds}.
The total uncertainty in the prediction of the top quark background is about 13\% (3\%)
in the zero-jet (one-jet) categories, and about 36\% in the $\PW+\text{jets}$ background
prediction. The total uncertainty in the $\dyll$ normalization is about 30\%,
including both statistical and systematic contributions. The uncertainties in
the yields of the $\dytt$,  $\PW\gamma$, and $\PW\gamma^{*}$ background processes
are 10\%, 30\%, and 40\%, respectively.

The theoretical uncertainties in the diboson cross sections are calculated by
varying the renormalization and factorization scales using the \MCFM
6.4 program~\cite{MCFM}. The effects of variations in the PDFs and of the value
of $\alpha_{s}$ on the predicted cross section are derived by following the same
prescription as for the signal acceptance. Including the experimental uncertainties
gives a systematic uncertainty of around 10\% for WZ and ZZ processes. In the case
of $\PW\Pgg^{(*)}$ backgrounds, the variation in PDFs gives a systematic uncertainty
of 4\%. A summary of the relative uncertainties in the $\PWp\PWm$ cross section
measurement is given in Table~\ref{tab:systematics:1}, where the jet counting model
uncertainty includes the renormalization and factorization scales, and underlying
event uncertainties.

\begin{table}[htbp]
\centering
\topcaption{\label{tab:systematics:1}Relative uncertainties in the $\PWp\PWm$ cross section measurement.}
\begin{tabular}{@{}l@{}c@{}}
\hline
Source                & Uncertainty (\%) \\
\hline
Statistical uncertainty           & 1.5              \\
\hline
Lepton efficiency                 & 3.8              \\
Lepton momentum scale             & 0.5              \\
Jet energy scale                  & 1.7              \\
$\MET$ resolution                 & 0.7              \\
$\ttbar$+$\cPqt\PW$ normalization & 2.2              \\
\PW+jets normalization   & 1.3              \\
$\dyll$ normalization             & 0.6              \\
$\dytt$ normalization             & 0.2              \\
$\PW\Pgg$ normalization           & 0.3              \\
$\PW\Pgg^{*}$ normalization       & 0.4              \\
$\V\V$ normalization              & 3.0              \\
$\PH\to\PWp\PWm$ normalization        & 0.8              \\
Jet counting theory model         & 4.3              \\
PDFs                              & 1.2              \\
MC statistical uncertainty                     & 0.9              \\
Integrated luminosity                        & 2.6              \\
\hline
Total uncertainty                 & 7.9    \\
\hline
\end{tabular}
\end{table}

\section{The \texorpdfstring{$\PWp\PWm$}{W+W-} cross section measurement}\label{sec:WWxsec}
The inclusive cross section is determined as

\begin{equation}
\sigma_{\PWp\PWm} = {\frac{N_\text{data}-N_\text{bkg} }{\mathcal{L} \, \epsilon \, \bigl( 3 \, \mathcal{B}(\PW\to \ell\cPagn) \bigr)^2}},
\label{eq:WWxsec:wwxs}
\end{equation}

where $N_\text{data}$ and $N_\text{bkg}$ are the total number of data and background events, $\epsilon$ is the signal efficiency, $\mathcal{L}$ is the integrated luminosity, and $\mathcal{B}(\PW\to \ell\cPagn)$ is the branching fraction for a $\PW$ boson decaying to each lepton family $\mathcal{B}(\PW\to \ell\cPagn) = (10.80 \pm 0.09)\%$~\cite{pdg}.

The signal efficiency $\epsilon$ is evaluated as the fraction of the sum of $\cPq\cPaq\to\PWp\PWm$ and $\Pg\Pg\to\PWp\PWm$ generated events, with $\PW\to \ell \nu$ ($\ell = \Pe$, $\mu$, $\tau$), accepted by the analysis selection. The efficiency estimated for each category is listed in Table~\ref{tbl:WWxsec:2}. The reported statistical uncertainty in the efficiency  originates from the limited size of the MC samples.

\begin{table*}[htpb]
\centering
\topcaption{\label{tbl:WWxsec:2}Signal efficiency for the four event categories used in the $\ppww$ cross section measurement. The values reported are a product of the detector geometrical acceptance and the object reconstruction and event identification efficiency. The statistical uncertainty is from the limited size of the MC samples.}
\begin{tabular}{llc}\hline
\multicolumn{2}{c}{Event category} & Signal efficiency (\%)\\\hline
\multirow{2}{*}{zero-jet category} & Different-flavor & $3.02 \pm 0.02\stat \pm 0.22\syst$\\
                                & Same-flavor      & $1.21 \pm 0.01\stat \pm 0.09\syst$\\
\multirow{2}{*}{one-jet category} & Different-flavor & $0.96 \pm 0.01\stat \pm 0.11\syst$\\
                                & Same-flavor      & $0.34 \pm 0.01\stat \pm 0.04\syst$\\\hline
\end{tabular}
\end{table*}

The $\WW$ production cross section in $\Pp\Pp$ collision data at $\sqrt{s}=8\TeV$ is measured separately in events with same- and different-flavor leptons and in events with exclusively zero or one reconstructed and identified jet, as shown in Table~\ref{tbl:WWxsec:3}. The number of events in each
category, as shown in Table~\ref{tbl:WWxsec:1}, is modeled as a Poisson random variable, whose mean
value is the sum of the contributions from the processes under consideration. Systematic uncertainties
are represented by individual nuisance parameters with log-normal distributions. The experimental and theoretical uncertainties in the event selection as well as the uncertainty on the integrated luminosity are reported separately. The theoretical component includes contributions from the jet counting theory model and PDFs as in Table~\ref{tab:systematics:1}.
The measurement in the different flavor final state is consistent with that in the same flavor final state at the level of 1.5$\sigma$ after
taking into account the statistical uncertainty and the uncorrelated systematic uncertainties.

\begin{table*}[htpb]
\centering
\topcaption{\label{tbl:WWxsec:3} The $\WW$ production cross section in each of the four event categories.}
\begin{tabular}{llc}\hline
\multicolumn{2}{c}{Event category} & $\WW$ production cross section (pb)\\\hline
\multirow{2}{*}{zero-jet category} & Different-flavor & $59.7\pm 1.1\stat\pm 3.3\experr\pm 3.5\thy\pm 1.6\lum$\\
                                & Same-flavor      & $64.3\pm 2.1\stat\pm 4.6\experr\pm 4.3\thy\pm 1.7\lum$\\
\multirow{2}{*}{one-jet category} & Different-flavor & $59.1\pm 2.8\stat\pm 6.0\experr\pm 6.2\thy\pm 1.6\lum$\\
                                & Same-flavor      & $65.1\pm 5.5\stat\pm 8.3\experr\pm 8.0\thy\pm 1.7\lum$\\\hline
\end{tabular}
\end{table*}

The four event categories are combined by performing a profile likelihood fit to the data following
the statistical methodology described in Refs.~\cite{LHC-HCG-Report,Read1,junkcls}. The combined result is:

\ifthenelse{\boolean{cms@external}}{
\begin{multline}\label{eq:WWxsec:1}
\sigma_{\PWp\PWm} = 60.1\pm 0.9\stat\pm 3.2\experr\pm 3.1\thy\\
\pm 1.6\lum\unit{pb} = 60.1\pm 4.8\unit{pb}.
\end{multline}
}{
\begin{equation}\label{eq:WWxsec:1}
\sigma_{\PWp\PWm} = 60.1\pm 0.9\stat\pm 3.2\experr\pm 3.1\thy\pm 1.6\lum\unit{pb} = 60.1\pm 4.8\unit{pb}.
\end{equation}
}

The combined result shows good agreement with the NNLO theoretical prediction
of $59.8^{+1.3}_{-1.1}\unit{pb}$~\cite{NNLOXsec}. The measurement precision
is dominated by the result in the different-flavor zero-jet event category. The
main source of systematic uncertainty comes from the modeling of the signal
efficiency, especially the requirement on the number of
reconstructed and identified jets.

We report the $\WW$ production cross section in a fiducial region defined by a jet veto requirement in order to be less sensitive to theoretical uncertainties related to the modelling of the signal efficiency, especially those related to the requirement on the number of reconstructed and identified jets. When specifying the fiducial regions at generation level, jets are defined at particle level, before the
detector effects, and clustered using the same anti-\kt algorithm with distance parameter of 0.5 as is used for collider data reconstruction. We measure the cross sections in a fiducial region defined 
by requiring no jets with  $\abs{\eta^{\text{jet}}}<4.7$ and jet \pt above a series of thresholds.  The results are summarized in Table~\ref{tbl:WWxsec:4} and compared with the predicted cross sections estimated with \POWHEG. These results are consistent with the SM expectations.

\begin{table*}[htbp]
\centering
\topcaption{\label{tbl:WWxsec:4}The $\WW$ production cross section in fiducial regions defined by requiring no jets at particle level with jet \pt thresholds as listed.}
{\begin{tabular}{ccc}\hline
$\pt^{\text{jet}}$ (\GeVns{}) & $\sigma_{\text{zero-jet}}$ measured (pb) &  $\sigma_{\text{zero-jet}}$ predicted  (pb) \\
\hline
$>$20       & $36.2\pm  0.6\stat \pm 2.1\experr\pm 1.1\thy\pm  0.9\lum$  &  $36.7\pm 0.1\stat$  \\
$>$25       & $40.8\pm  0.7\stat \pm 2.3\experr\pm 1.3\thy\pm  1.1\lum$  &  $40.9\pm 0.1\stat$  \\
$>$30       & $44.0\pm  0.7\stat \pm 2.5\experr\pm 1.4\thy\pm  1.1\lum$  &  $43.9\pm 0.1\stat$   \\
\hline
\end{tabular}
}
\end{table*}

The $\WW$ cross section is also measured in the different-flavor zero-jet category, which is the most precise channel.
The fiducial region is defined at
generation level by requiring no jets with $\abs{\eta^{\text{jet}}}<4.7$ and a given maximum jet \pt for events with prompt leptons with $\pt>20\GeV$ and $\abs{\eta}<2.5$ before final-state radiation. In this case leptonic $\tau$
decays are not considered as part of the signal. The signal efficiency for this selection at generator level
excluding $\tau$ lepton decays is 31.8\% for a jet \pt threshold of $30\GeV$. The measured cross sections are summarized in Table~\ref{tbl:WWxsec:5} and compared with the predicted cross sections estimated with \POWHEG.

\begin{table*}[htbp]
\centering
\topcaption{\label{tbl:WWxsec:5}The $\WW$ production cross section in fiducial regions defined by requiring zero jets at particle level with varying jet \pt thresholds and requiring prompt leptons with $\pt>20\GeV$ and $\abs{\eta}<2.5$, before final-state radiation. }
{\begin{tabular}{ccc}\hline
$\pt^{\text{jet}}$ (\GeVns{}) & $\sigma_{{\text{zero-jet}},\PW\to\ell\nu}$ measured (fb) &  $\sigma_{{\rm{zero-jet}},\PW\to\ell\nu}$ predicted (fb) \\
\hline
$>$20        &  $ 223\pm  4\stat\pm 13\experr\pm 7\thy\pm  6\lum$ &  $ 228\pm 1\stat$ \\
$>$25        &  $ 253\pm  5\stat\pm 14\experr\pm 8\thy\pm  7\lum$ &  $ 254\pm 1\stat$ \\
$>$30        &  $ 273\pm  5\stat\pm 15\experr\pm 9\thy\pm  7\lum$ &  $ 274\pm 1\stat$ \\
\hline
\end{tabular}
}
\end{table*}

Since both fiducial cross section measurements are restricted to the zero-jet category, most systematic uncertainties are calculated in the same way as in the inclusive analysis, except the underlying event, PDFs, and renormalization and factorization scales effects related to the $\PWp\PWm$ signal. In these cases the uncertainty is estimated as the largest difference among the three signal MC generators, \POWHEG, \MADGRAPH, and \MCATNLO, for the fraction of reconstructed events outside the fiducial region and passing the full analysis selection. Fractionally, the theoretical uncertainty changes from 5\% to 3\%.

\section{Normalized differential \texorpdfstring{$\PWp\PWm$}{W+W} cross section measurement}\label{sec:diffWWxsec}
{\tolerance=80
The normalized differential $\WW$ cross section\breakhere
${(1/\sigma)} \, {\rd\sigma}/{\rd X}$ is determined as a function of
different
$X$ variables: the leading lepton $p_{{\mathrm{T}},\text{ max}}^{\ell}$, the
transverse momentum of the dilepton system
$\pt^{\ell\ell}$, the invariant mass $\mll$, and the angular separation in the transverse plane
between the two leptons $\Delta\phi_{\ell\ell}$. The measurements are performed using unfolded
distributions from events with zero jets and the $\Pepm\PGmmp$ final state only. Leptonic $\tau$ decays are
not considered as part of the signal.
\par}

The fiducial cross section is determined by the event yield in each bin after subtracting backgrounds. Each distribution is then
corrected for event selection efficiencies and for detector resolution
effects in order to be compared with predictions from event generators. The detector resolution corrections vary between 5\% and 15\% depending on the variable and the bin. The correction procedure is based on unfolding techniques, as implemented in the RooUnfold toolkit~\cite{arxiv:1105.1160}, which provides both singular value decomposition (SVD)~\cite{nima:a372_469_481} and the
iterative Bayesian~\cite{nima:a362_487_498} methods. Both algorithms use a response matrix that correlates the observable with and without detector effects.
Regularization parameters are tuned to obtain results that are robust
against numerical instabilities and statistical fluctuations. The unfolding is
performed with the SVD method, and we cross-check the results with the iterative
Bayesian method. We found a good agreement within uncertainties between both methods.
The differential cross section is derived by dividing the corrected number of events by the
integrated luminosity and by the bin width.

For each measured distribution, a response matrix is evaluated using $\cPq\cPaq\to\PWp\PWm$ events
(generated with \POWHEG) and $\Pg\Pg\to\PWp\PWm$, after full detector simulation. In order to minimize the model uncertainties
due to unnecessary extrapolations of the measurement outside the experimentally well-described
phase space region, the normalized differential cross section is determined in a phase
space defined at the particle level by considering prompt leptons before final-state radiation,
with $\pt > 20\GeV$ and $\abs{\eta} < 2.5$. Events with one or more jets with
$\pt >30$\GeV and $\abs{\eta}<$ 4.7 are rejected.

The systematic uncertainties in each bin are assessed from the variations of the
nominal cross section by repeating the full analysis for every systematic
variation. The difference with respect to the nominal value is taken as the
final systematic uncertainty for each bin and each measured observable. By using
this method, the possible correlations of the systematic uncertainties between
bins are taken into account. Those systematic uncertainties that are correlated across all bins of the measurement, and
therefore mainly affect the normalization, cancel out at least partially in the normalized cross section. The
uncertainty also includes the statistical error propagation through the unfolding
method using the covariance matrix and the difference in the response matrix
from \MADGRAPH, \POWHEG, and \MCATNLO, the latter being almost negligible.

\begin{figure*}[phtb]
  \begin{center}
    \includegraphics[width=0.45\textwidth]{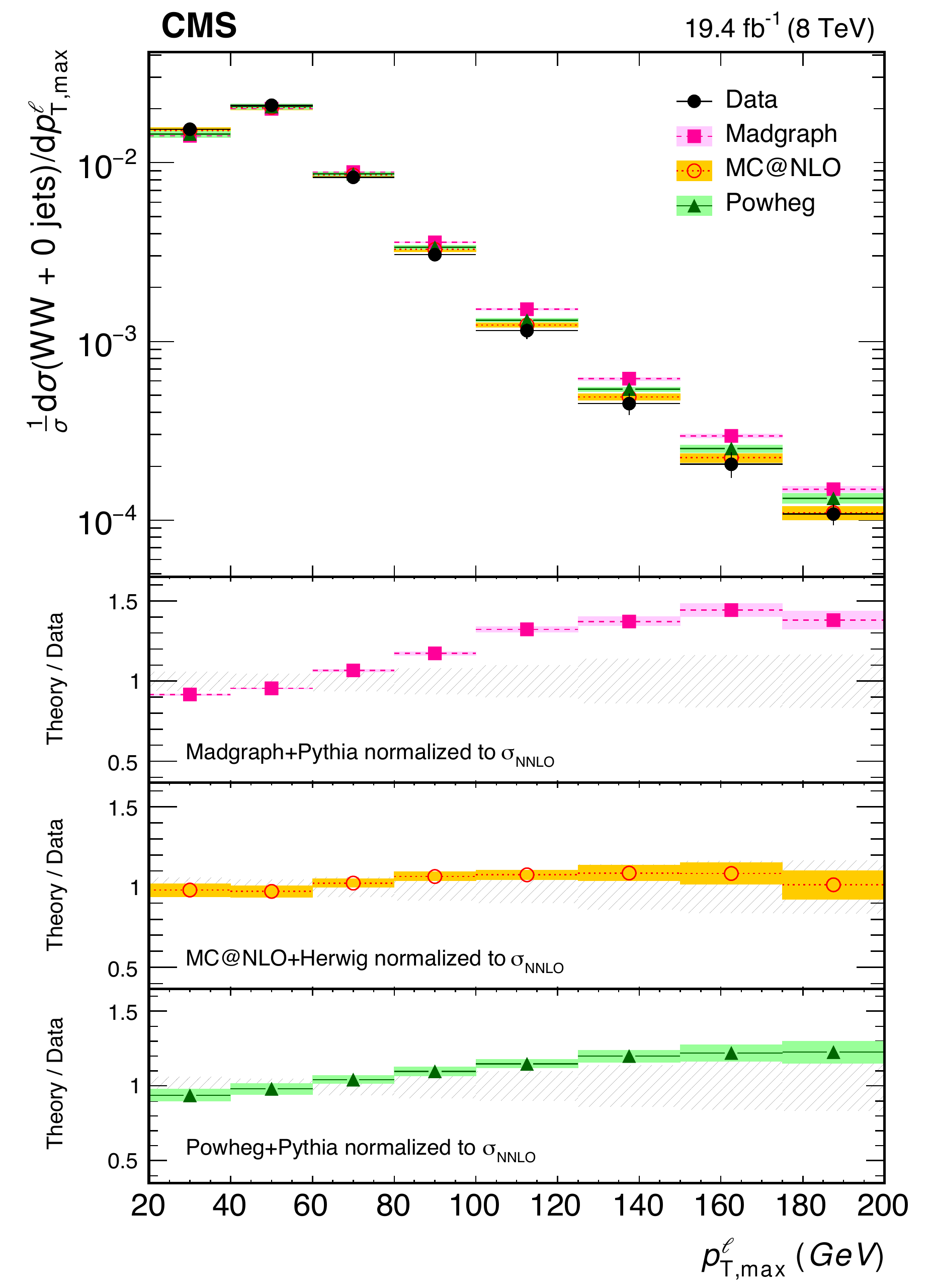}
    \includegraphics[width=0.45\textwidth]{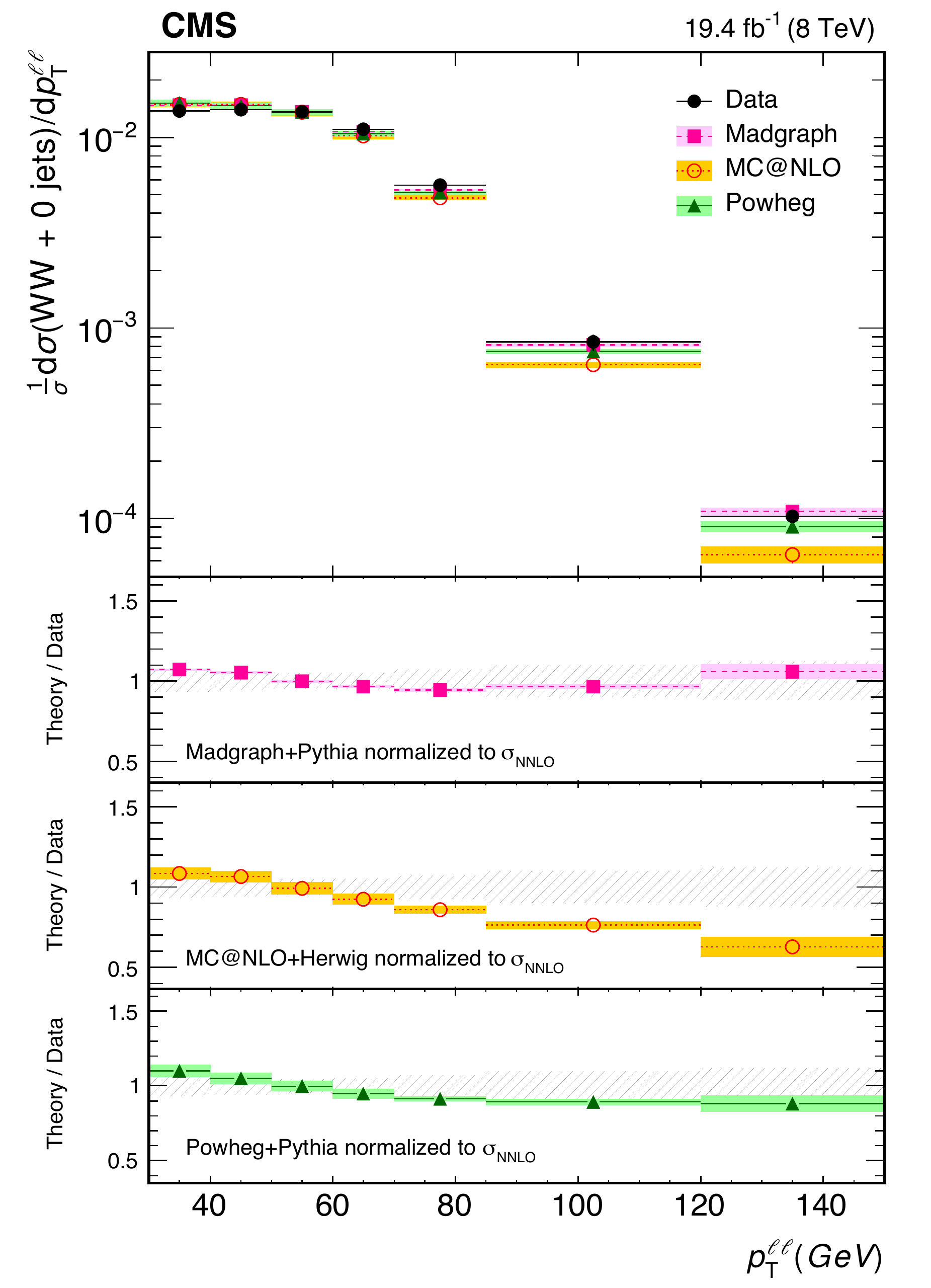}\vspace*{0.3cm}
    \includegraphics[width=0.45\textwidth]{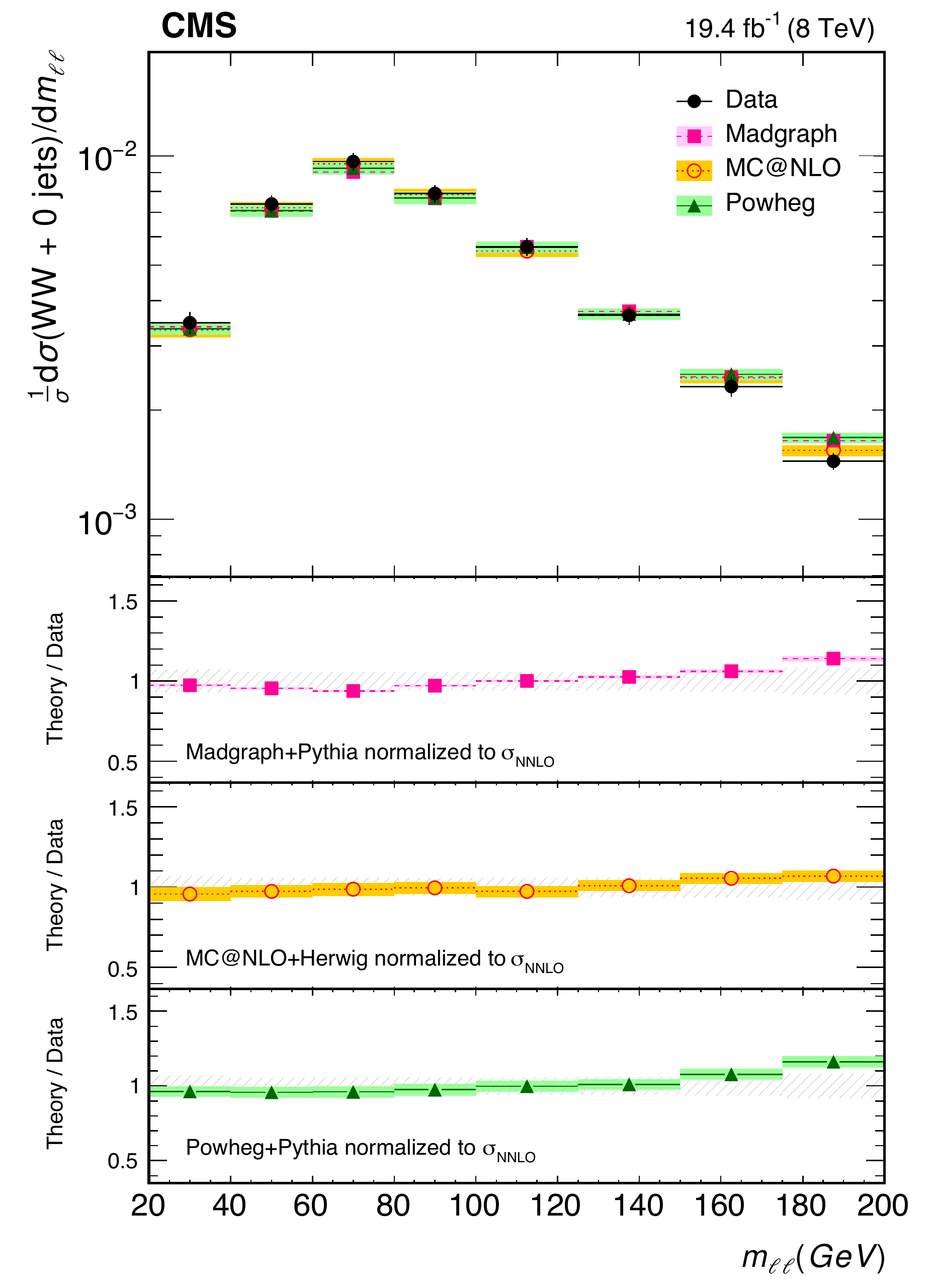}
    \includegraphics[width=0.45\textwidth]{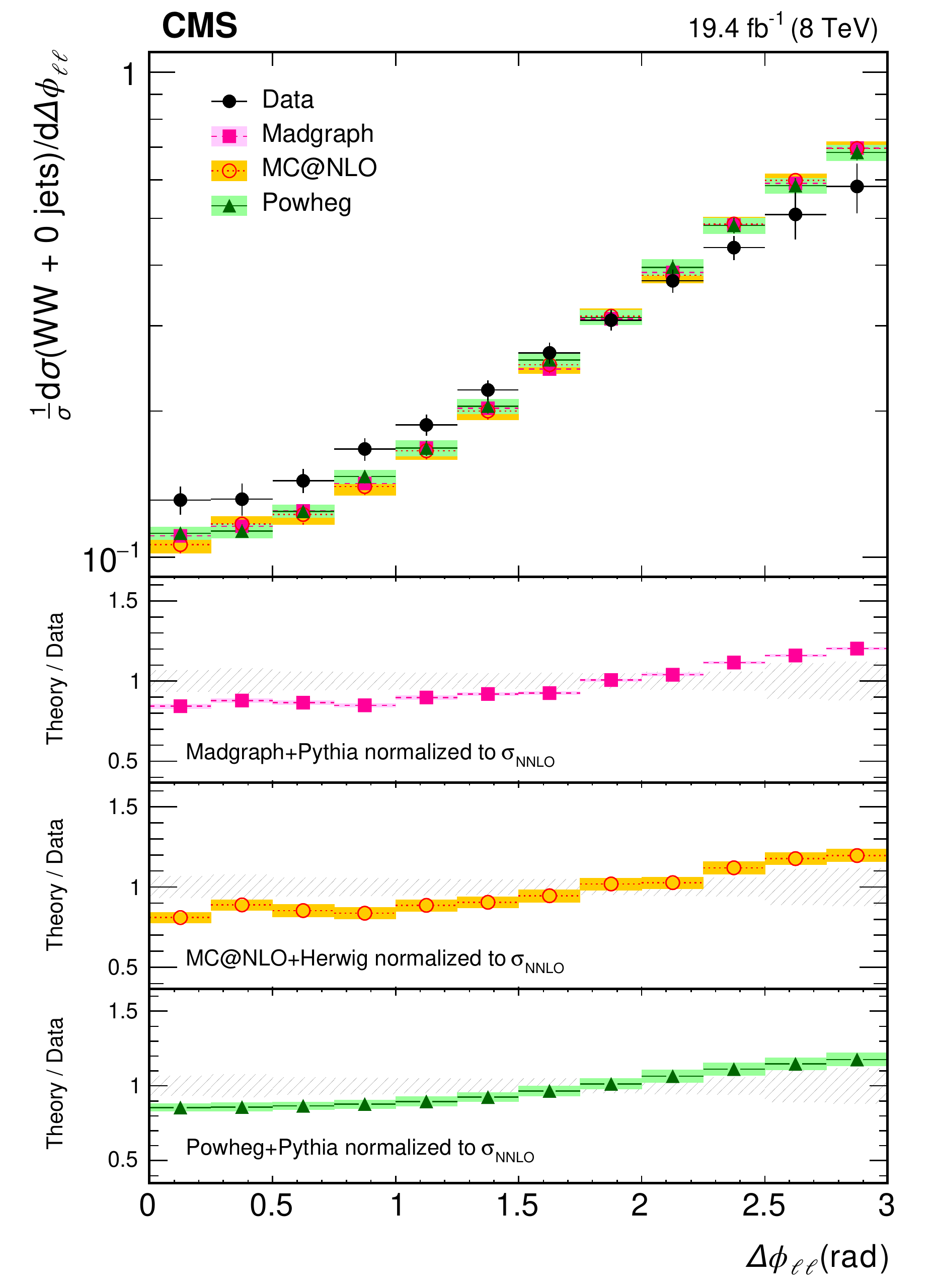}\vspace*{0.3cm}
    \caption{\label{fig:diffXSec} Normalized differential $\WW$
    cross section as a function of the leading lepton \pt ($p_{{\mathrm{T}},\text{ max}}^{\ell}$) (top left), the
    transverse momentum of the dilepton system ($\pt^{\ell\ell}$) (top right),
    the invariant mass ($\mll$) (bottom left) and  the angular separation between
    leptons ($\Delta\phi_{\ell\ell}$) (bottom right). Both statistical and systematic uncertainties
    are included. The hatched area in the ratio plots corresponds to the relative error of the data in each bin. The measurement, including $\Pg\Pg\to\PWp\PWm$ is compared to predictions from \MADGRAPH, \POWHEG, and \MCATNLO.}
  \end{center}
\end{figure*}

Various differential cross sections in interesting kinematic variables are
presented in Fig.~\ref{fig:diffXSec}. The measurements, including $\Pg\Pg\to\PWp\PWm$,
are compared to the predictions from \MADGRAPH, \POWHEG, and
\MCATNLO, normalized to the recent QCD calculations up to approximate
NNLO precision~\cite{NNLOXsec}. The predictions from \MADGRAPH are shown with
statistical uncertainties only. No single generator performs best for all the kinematic variables, although
\POWHEG does better than the others. Data and theory show a good agreement for the $\mll$ and the $\pt^{\ell\ell}$ distributions, within uncertainties, except for the \MCATNLO generator which predicts a softer $\pt^{\ell\ell}$ spectrum than observed. In case of the $p_{{\mathrm{T}},\text{ max}}^{\ell}$ distribution, the
\MADGRAPH prediction shows an excess of events in the tail of the distribution compared to data,
while \POWHEG shows a reasonable agreement and \MCATNLO shows a good agreement. We observe more significant differences in the shape of the $\Delta\phi_{\ell\ell}$ for all the three generators as compared to the data. Depending on the choice of MC generator, some of the differential cross sections show discrepancies up to 20\%, in extreme cases even up to 50\%, when comparing with a LO generator. These deviations are covered by the typical background uncertainties of Run 1 searches for physics beyond the SM. A better modelling of the WW background will be required to reduce the corresponding systematic uncertainties for Run 2, however.

\section{Limits on anomalous gauge couplings}\label{sec:aTGC}
{\tolerance=800
Beyond-standard-model (BSM) physics effects in $\ppww$ can be described by a series
of operators with mass dimensions larger than four in addition to the dimension-four
operators in the SM Lagrangian.
In the electroweak sector of the SM, in an EFT interpretation~\cite{eftbasis}, the
first higher-dimension operators made solely from electroweak vector fields and the
Higgs doublet have mass dimension six. There are six different dimension-six operators
that generate ATGCs. Three of them are C- and P-conserving while the others are not.
In this analysis, we only consider models with C- and P-conserving operators. In the HISZ basis~\cite{HISZ},
these three operators are written as:
\begin{align}\label{atgcs:eq:1}
\begin{split}
\frac{c_{\mathrm{WWW}}}{\Lambda^2}\mathcal{O}_{\mathrm{WWW}} &= \frac{c_{\mathrm{WWW}}}{\Lambda^2}\mathrm{Tr}[{\mathrm{W}}_{\mu\nu}{\mathrm{W}}^{\nu\rho}W_{\rho}^{\;\mu}],\\
\frac{c_{\mathrm{W}}}{\Lambda^2}\mathcal{O}_{\mathrm{W}} &= \frac{c_{\mathrm{W}}}{\Lambda^2}(D^{\mu}\Phi)^{\dagger}{\mathrm{W}}_{\mu\nu}(D^{\nu}\Phi),\\
\frac{c_{\mathrm{B}}}{\Lambda^2}\mathcal{O}_{\rm{B}} &= \frac{c_{\mathrm{B}}}{\Lambda^2}(D^{\mu}\Phi)^{\dagger}{\mathrm{B}}_{\mu\nu}(D^{\nu}\Phi).\\
\end{split}
\end{align}

The parameter $\Lambda$ is the mass scale that characterizes the coefficients of
the higher-dimension operators, which can be regarded as the scale of new physics.
The three operators in Eq.~\eqref{atgcs:eq:1} generate both ATGC and Higgs
boson anomalous couplings at tree level and modify the $\ppww$ cross section.
In the absence of momentum-dependent form factors, the traditional LEP
parametrization of ATGCs can be related to the values of the coupling
constants of the dimension-six electroweak operators~\cite{eftbasis} as summarized in Eq.~\ref{eq:basisconv}:

\begin{equation}\label{eq:basisconv}
\begin{split}
&\delta(c_{\text{WWW}}/\Lambda^2) = \frac{2}{3g^2M_{\PW^2}}\delta \lambda_{\gamma},\\
&\delta(c_{\text{W}}/\Lambda^2) = \frac{2}{M_{\Z^2}}\delta g_1^\Z,\\
&\delta(c_{\text{B}}/\Lambda^2) = 2\sqrt{\left(\frac{\delta \kappa_{\gamma}}{M_{\PW^2}}\right)^2+\left(\frac{\delta g_1^\Z}{M_{\Z^2}}\right)^2.}
\end{split}
\end{equation}

The dataset selected for the $\PWp\PWm$ cross section measurement is used to bound $c_{{\mathrm{WWW}}}/\Lambda^2$, $c_{\mathrm{W}}/\Lambda^2$, and $c_{\mathrm{B}}/\Lambda^2$.
For this measurement, we require the events to have zero reconstructed and identified
jets with $\pt > 30\GeV$ and $\abs{\eta} < 4.7$. We use the $m_{\ell\ell}$
distribution because it is robust against mismodeling of the transverse boost of the $\WW$
system and is sensitive to the value of the coupling constants associated with the
dimension-six operators. A binned Poisson log-likelihood comparing the data and simulated $m_{\ell\ell}$ distributions is computed.
The template histograms representing various values of the ATGCs  are prepared using $\PWp\PWm$
simulated events generated with \MADGRAPH using a Lagrangian that contains
the SM interaction terms and the three operators above. Thus, the simulation includes
the pure SM contribution, the ATGC contribution, the Higgs boson anomalous coupling
contribution, and the interference between the SM and ATGC contributions. The hard-scattering simulation
includes up to one hard parton in the final state~\cite{MadgraphMLM}. The detector response
to the events is obtained using the detailed CMS detector simulation.
The various background yields described in Section~\ref{sec:backgrounds} are
added to the $\mll$ distribution from the simulated signal events. As an
example of the templates, Fig.~\ref{atgcs:fig:1} shows the $m_{\ell\ell}$
distribution for one set of values of $c_{\mathrm{WWW}}/\Lambda^2$, $c_{\mathrm{W}}/\Lambda^2$, and $c_{\mathrm{B}}/\Lambda^2$.

\begin{figure}[htpb]
\centering
\includegraphics[width=0.48\textwidth]{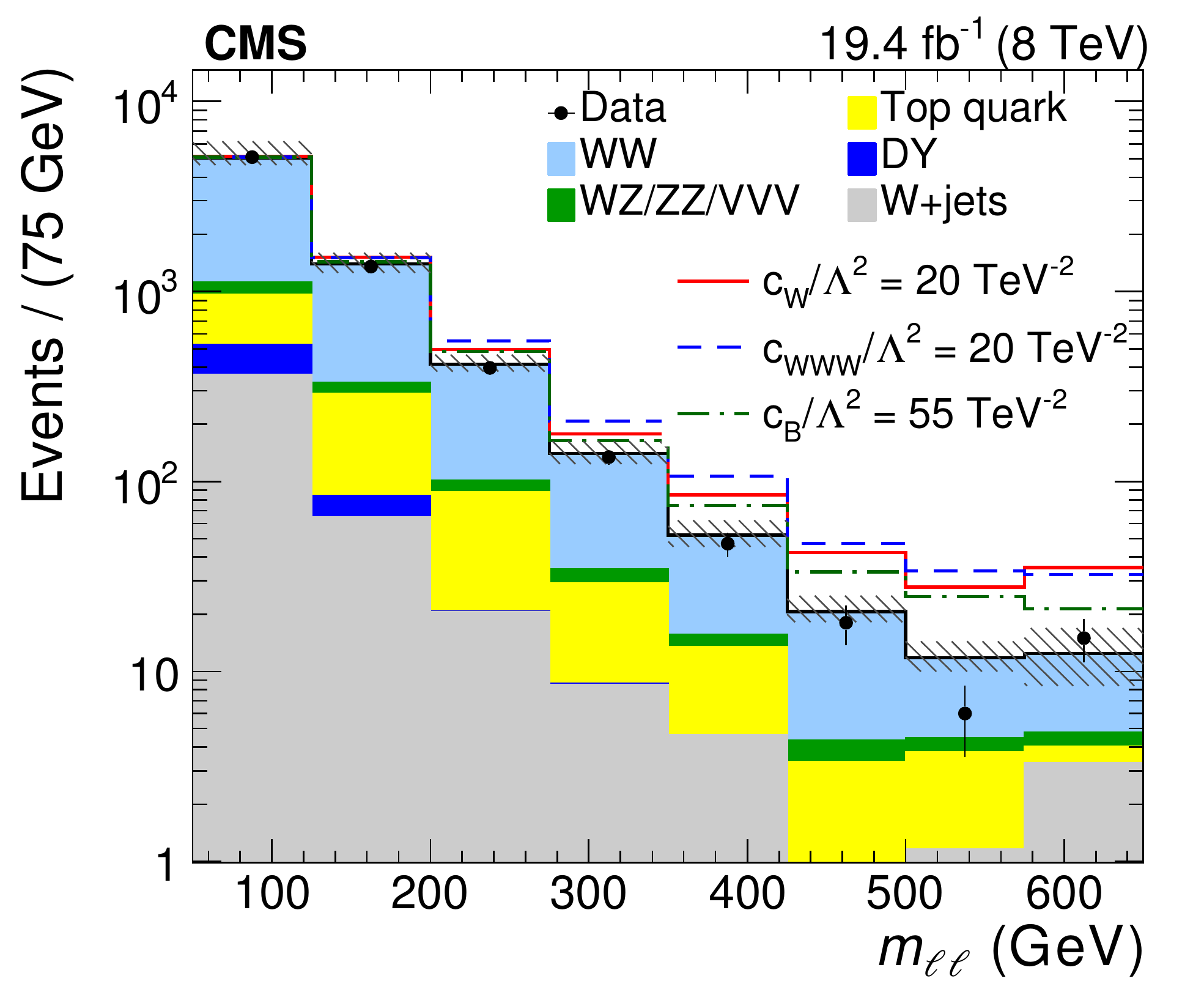}
\caption{\label{atgcs:fig:1}The $\mll$ distribution with
all SM backgrounds and $c_{\mathrm{W}}/\Lambda^2 = 20\TeV^{-2}$,
$c_{\mathrm{WWW}}/\Lambda^2 = 20\TeV^{-2}$, and
$c_{\mathrm{B}}/\Lambda^2 = 55\TeV^{-2}$. The events are selected requiring no
reconstructed jets with $\pt > 30\GeV$ and $\abs{\eta} < 4.7$. The last
bin includes all events with $\mll > 575\GeV$. The hatched area
around the SM distribution is the total systematic uncertainty in each bin.
The signal component is simulated with \MADGRAPH and contains the
$\cPq\cPaq\to\PWp\PWm$, the nonresonant $\cPg\cPg \to \PWp\PWm$, and the
$\cPg\cPg \to \PH\to\PWp\PWm$ components.}
\end{figure}

\begin{figure}[htpb]
\centering
\includegraphics[width=0.48\textwidth]{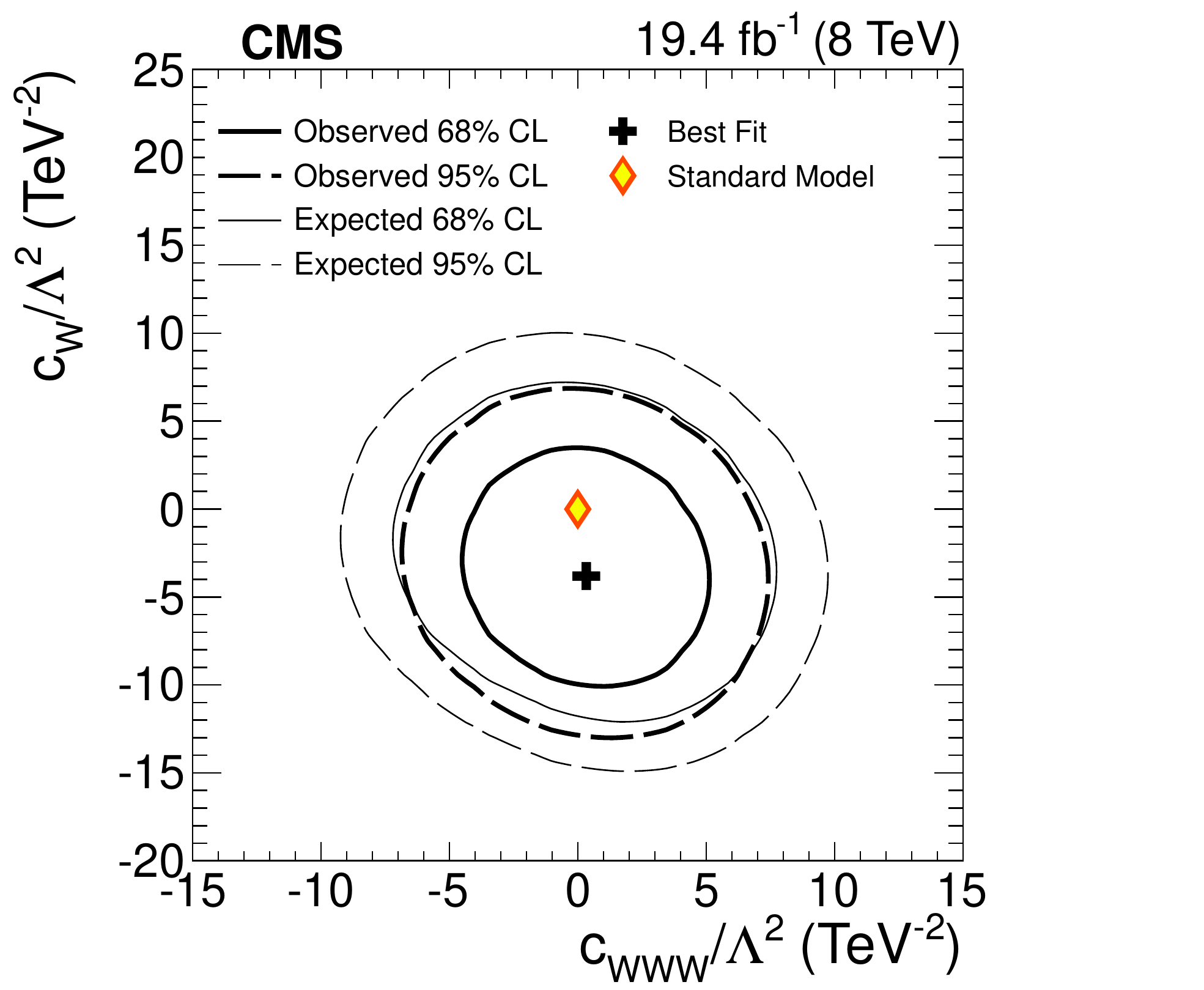}
\includegraphics[width=0.48\textwidth]{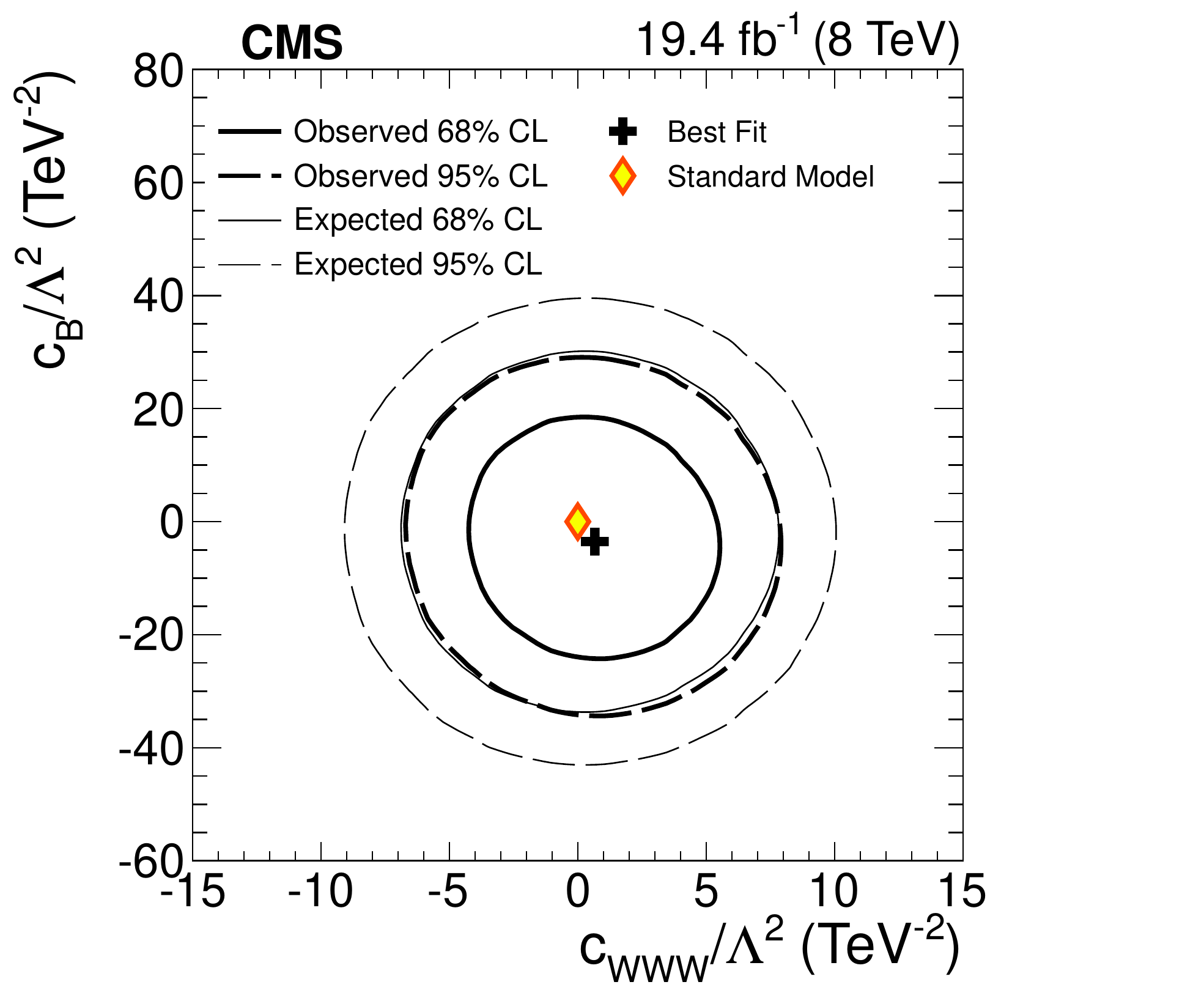}
\includegraphics[width=0.48\textwidth]{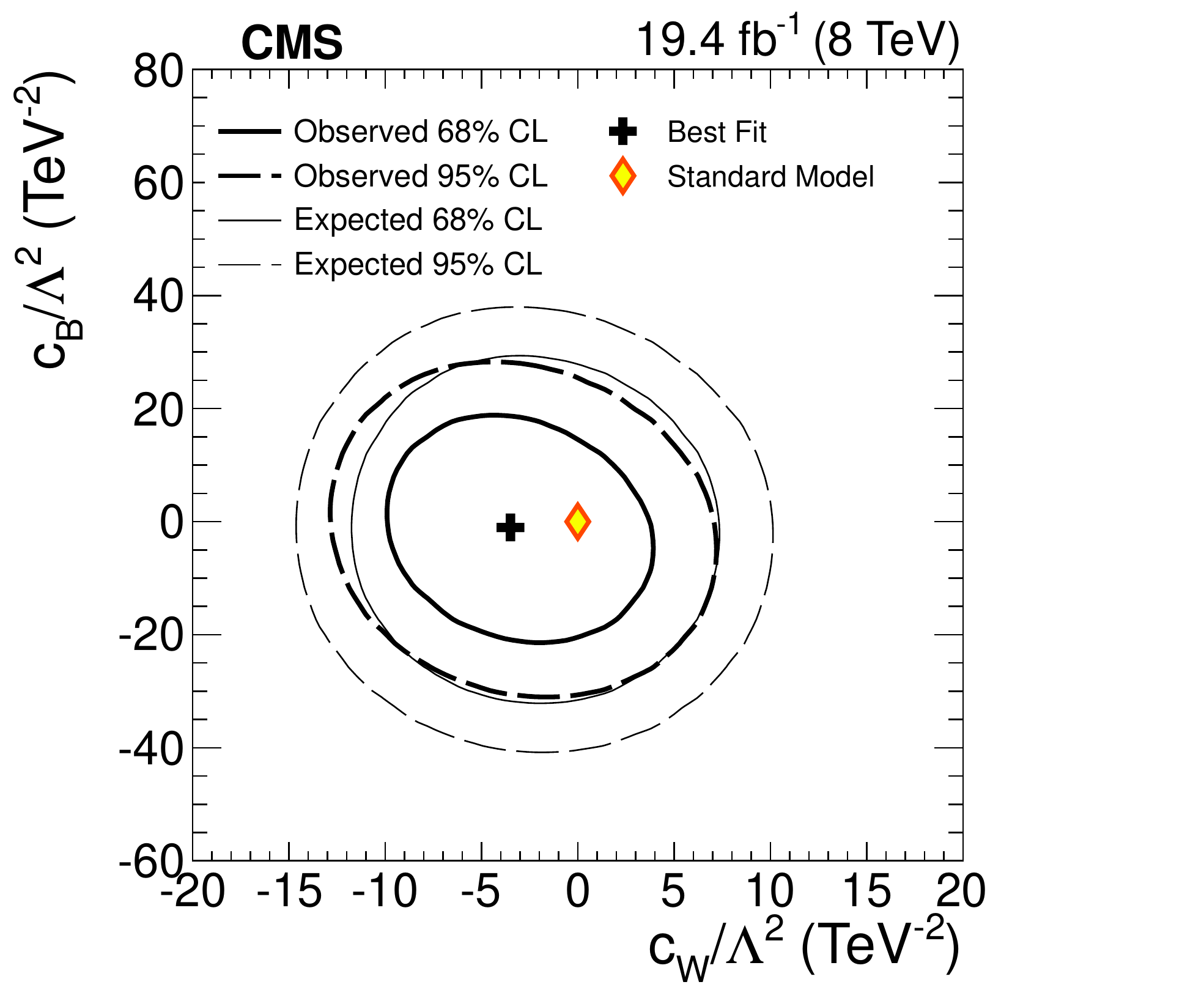}
\caption{\label{atgcs:fig:2}Two-dimensional observed (thick lines) and expected (thin lines) 68\% and 95\% CL contours. The contours are obtained from profile log-likelihood comparisons to data assuming two nonzero coupling constants: $c_{\mathrm{WWW}}/\Lambda^2 \times c_{\mathrm{W}}/\Lambda^2$, $c_{\mathrm{WWW}}/\Lambda^2 \times c_{\mathrm{B}}/\Lambda^2$, and $c_{\mathrm{W}}/\Lambda^2 \times c_{\mathrm{B}}/\Lambda^2$. The cross markers indicate the best-fit values, and the diamond markers indicate the SM ones.}
\end{figure}

{\tolerance=800
Templates of the  $\mll$  distribution are prepared for different hypothetical
values of the coupling constants $c_{\mathrm{WWW}}/\Lambda^2$, $c_{\mathrm{W}}/\Lambda^2$, and
$c_{\mathrm{B}}/\Lambda^2$. We consider both the cases in which only one of the
coupling constants has a nonzero value, and the cases in which two of them
are varied simultaneously. The correlations between the measured coupling constants
are not strong, so we do not consider the case in which the three
coupling constants are allowed to vary simultaneously. Thus, the results presented
here assume that the symmetries of the BSM theory would only allow either one or two of
the dimension-six electroweak operators to contribute appreciably.
\par}

The expected number of events in each bin of the template histograms is interpolated
using polynomial functions as a function of the coupling constants to create
a continuous parametrization of the model. A profile likelihood fit to the data
for each coupling-constant hypothesis is performed using the  method
described in Section~\ref{sec:WWxsec}.

\begin{table*}[htbp]
\topcaption{\label{tbl:results} Measured $c_{\mathrm{WWW}}/\Lambda^2$, $c_{\mathrm{W}}/\Lambda^2$, and $c_{\mathrm{B}}/\Lambda^2$ coupling
constants and their corresponding 95\% CL intervals. Results are compared to the world average values, as explained in the text.}
\begin{center}
\begin{tabular}{l c c r l}
\hline
Coupling constant    & This result            & Its 95\% CL interval  & \multicolumn{2}{c}{World average} \\
                     & ($\TeVns^{-2}$)    & ($\TeVns^{-2}$)        & \multicolumn{2}{c}{($\TeVns^{-2}$)} \\
\hline
$c_{\mathrm{WWW}}/\Lambda^2$  &  $0.1_{-3.2}^{+3.2}$     & [$-5.7,  5.9$]           & $-5.5\pm 4.8$ & (from $\lambda_{\gamma}$) \\
$c_{\mathrm{W}}/\Lambda^2$    &  $-3.6_{-4.5}^{+5.0}$    & [$-11.4, 5.4$]           & $-3.9^{+3.9}_{-4.8}$ & (from $g_1^{\Z}$)       \\
$c_{\mathrm{B}}/\Lambda^2$    &  $-3.2_{-14.5}^{+15.0}$  & [$-29.2, 23.9$]          & $-1.7^{+13.6}_{-13.9}$ & (from $\kappa_{\gamma}$ and $g_1^\Z$)\\
\hline
\end{tabular}
\end{center}
\end{table*}

Figure~\ref{atgcs:fig:2} shows the 2D likelihood profiles at 68\% and 95\% confidence levels (CL) for the three cases
in which two coupling constants are allowed to vary. Using the templates prepared with a single non-zero coupling constant,
we measure the values of $c_{\mathrm{WWW}}/\Lambda^2$, $c_{\mathrm{W}}/\Lambda^2$, and $c_{\mathrm{B}}/\Lambda^2$ individually.
The result of the 1D likelihood fit at 95\% CL intervals are given in Table~\ref{tbl:results}.

In general, EFT predictions are valid if they maintain a separation between the scale of the momentum transfer in the process and the scale of new physics and if they preserve unitarity~\cite{Biekoetter:2014jwa}. The first condition implies an upper bound on $|(c/\Lambda^2)\SHat|$ of $(4\pi)^2 \approx 158$, although a specific new physics model may be more restrictive. The second condition requires an analysis of each operator, and sets the limits~\cite{Corbett:2014ora}: $|(c_{\text{WWW}}/\Lambda^2)\SHat| < 85$, $|(c_{\text{W}}/\Lambda^2)\SHat| < 205$, and $|(c_{\text{B}}/\Lambda^2)\SHat| < 640$. For the experimental limits on the operator $\mathcal{O}_{\text{WWW}}$ given on Table~\ref{tbl:results}, the most stringent constraint comes from the second condition and implies validity for $\rootSHat < 3.8\TeV$. The operators $\mathcal{O}_{\text{W}}$ and $\mathcal{O}_{\text{B}}$ are constrained by the first condition to be valid for $\rootSHat < 3.7\TeV$ and $\rootSHat < 2.3\TeV$, respectively. In all three cases we expect all the data to have $\rootSHat$ within the EFT range of validity. At the extreme hypothesis, for which the bounds are derived, only 3\% of the selected $\WW$ events are expected to have $\rootSHat > 2.3\TeV$. Within the limits of this interpretation, no evidence for anomalous WWZ and WW$\gamma$ triple gauge-boson couplings is found. Our results are compared to the world average values expressed in terms of $\lambda_{\gamma}$, $g_1^\Z$ and $\kappa_{\gamma}$ couplings. These world average values are driven by the LEP results~\cite{Schael:2013ita,pdg}. The conversion of the world average values from $\lambda_{\gamma}$, $g_1^\Z$ and $\kappa_{\gamma}$ couplings to the EFT formalism is done using the results from Ref.~\cite{eftbasis} and ignoring correlations as summarized in Eq.~\ref{eq:basisconv}. These results represent an
improvement in the measurement of $c_{\text{WWW}}/\Lambda^{2}$.

\section{Summary}\label{sec:conclusions}
This paper reports a measurement of the $\PWp\PWm$  cross section
in pp collisions at a center of mass energy of 8\TeV, using an integrated luminosity of
$\mathcal{L} = 19.4\pm 0.5\fbinv$.  The measured $\PWp\PWm$
cross section is $60.1 \pm 0.9\stat\pm 3.2\experr\pm 3.1\thy\pm 1.6\lum\unit{pb}$
= $60.1\pm 4.8\unit{pb}$, consistent with the NNLO theoretical prediction
$\sigma^{\mathrm{NNLO}} (\Pp\Pp \to \WW) = 59.8^{+1.3}_{-1.1}\unit{pb}$.
We also report results on the normalized differential cross section measured as a function of kinematic
variables of the final-state charged leptons and compared with several predictions from perturbative
QCD calculations.
Data and theory show a good agreement for the $\mll$ and the $\pt^{\ell\ell}$ distributions
within uncertainties, but the \MCATNLO generator predicts a softer $\pt^{\ell\ell}$ spectrum
compared with the data events. In case of the $p_{{\mathrm{T}},\text{ max}}^{\ell}$ distribution, the
\MADGRAPH prediction shows an excess of events in the tail of the distribution compared to data,
while \POWHEG shows a reasonable agreement and \MCATNLO shows a good agreement.
We also observed differences in the shape of the $\Delta\phi_{\ell\ell}$ for the three
generators compared to the data.
No evidence for anomalous WWZ and WW$\gamma$ triple gauge-boson couplings is found,
and limits on their magnitudes are set. These new limits are comparable to the current world average,
and represent an improvement in the measurement of the coupling constant $c_{\mathrm{WWW}}/\Lambda^{2}$.

\begin{acknowledgments}\label{sec:Acknowledgements}
\hyphenation{Bundes-ministerium Forschungs-gemeinschaft Forschungs-zentren} We congratulate our colleagues in the CERN accelerator departments for the excellent performance of the LHC and thank the technical and administrative staffs at CERN and at other CMS institutes for their contributions to the success of the CMS effort. In addition, we gratefully acknowledge the computing centers and personnel of the Worldwide LHC Computing Grid for delivering so effectively the computing infrastructure essential to our analyses. Finally, we acknowledge the enduring support for the construction and operation of the LHC and the CMS detector provided by the following funding agencies: the Austrian Federal Ministry of Science, Research and Economy and the Austrian Science Fund; the Belgian Fonds de la Recherche Scientifique, and Fonds voor Wetenschappelijk Onderzoek; the Brazilian Funding Agencies (CNPq, CAPES, FAPERJ, and FAPESP); the Bulgarian Ministry of Education and Science; CERN; the Chinese Academy of Sciences, Ministry of Science and Technology, and National Natural Science Foundation of China; the Colombian Funding Agency (COLCIENCIAS); the Croatian Ministry of Science, Education and Sport, and the Croatian Science Foundation; the Research Promotion Foundation, Cyprus; the Ministry of Education and Research, Estonian Research Council via IUT23-4 and IUT23-6 and European Regional Development Fund, Estonia; the Academy of Finland, Finnish Ministry of Education and Culture, and Helsinki Institute of Physics; the Institut National de Physique Nucl\'eaire et de Physique des Particules~/~CNRS, and Commissariat \`a l'\'Energie Atomique et aux \'Energies Alternatives~/~CEA, France; the Bundesministerium f\"ur Bildung und Forschung, Deutsche Forschungsgemeinschaft, and Helmholtz-Gemeinschaft Deutscher Forschungszentren, Germany; the General Secretariat for Research and Technology, Greece; the National Scientific Research Foundation, and National Innovation Office, Hungary; the Department of Atomic Energy and the Department of Science and Technology, India; the Institute for Studies in Theoretical Physics and Mathematics, Iran; the Science Foundation, Ireland; the Istituto Nazionale di Fisica Nucleare, Italy; the Ministry of Science, ICT and Future Planning, and National Research Foundation (NRF), Republic of Korea; the Lithuanian Academy of Sciences; the Ministry of Education, and University of Malaya (Malaysia); the Mexican Funding Agencies (CINVESTAV, CONACYT, SEP, and UASLP-FAI); the Ministry of Business, Innovation and Employment, New Zealand; the Pakistan Atomic Energy Commission; the Ministry of Science and Higher Education and the National Science Centre, Poland; the Funda\c{c}\~ao para a Ci\^encia e a Tecnologia, Portugal; JINR, Dubna; the Ministry of Education and Science of the Russian Federation, the Federal Agency of Atomic Energy of the Russian Federation, Russian Academy of Sciences, and the Russian Foundation for Basic Research; the Ministry of Education, Science and Technological Development of Serbia; the Secretar\'{\i}a de Estado de Investigaci\'on, Desarrollo e Innovaci\'on and Programa Consolider-Ingenio 2010, Spain; the Swiss Funding Agencies (ETH Board, ETH Zurich, PSI, SNF, UniZH, Canton Zurich, and SER); the Ministry of Science and Technology, Taipei; the Thailand Center of Excellence in Physics, the Institute for the Promotion of Teaching Science and Technology of Thailand, Special Task Force for Activating Research and the National Science and Technology Development Agency of Thailand; the Scientific and Technical Research Council of Turkey, and Turkish Atomic Energy Authority; the National Academy of Sciences of Ukraine, and State Fund for Fundamental Researches, Ukraine; the Science and Technology Facilities Council, UK; the US Department of Energy, and the US National Science Foundation.

Individuals have received support from the Marie-Curie program and the European Research Council and EPLANET (European Union); the Leventis Foundation; the A. P. Sloan Foundation; the Alexander von Humboldt Foundation; the Belgian Federal Science Policy Office; the Fonds pour la Formation \`a la Recherche dans l'Industrie et dans l'Agriculture (FRIA-Belgium); the Agentschap voor Innovatie door Wetenschap en Technologie (IWT-Belgium); the Ministry of Education, Youth and Sports (MEYS) of the Czech Republic; the Council of Science and Industrial Research, India; the HOMING PLUS program of the Foundation for Polish Science, cofinanced from European Union, Regional Development Fund; the Compagnia di San Paolo (Torino); the Consorzio per la Fisica (Trieste); MIUR project 20108T4XTM (Italy); the Thalis and Aristeia programs cofinanced by EU-ESF and the Greek NSRF; the National Priorities Research Program by Qatar National Research Fund; the Rachadapisek Sompot Fund for Postdoctoral Fellowship, Chulalongkorn University (Thailand); and the Welch Foundation.
\end{acknowledgments}
\bibliography{auto_generated}

\cleardoublepage \appendix\section{The CMS Collaboration \label{app:collab}}\begin{sloppypar}\hyphenpenalty=5000\widowpenalty=500\clubpenalty=5000\textbf{Yerevan Physics Institute,  Yerevan,  Armenia}\\*[0pt]
V.~Khachatryan, A.M.~Sirunyan, A.~Tumasyan
\vskip\cmsinstskip
\textbf{Institut f\"{u}r Hochenergiephysik der OeAW,  Wien,  Austria}\\*[0pt]
W.~Adam, E.~Asilar, T.~Bergauer, J.~Brandstetter, E.~Brondolin, M.~Dragicevic, J.~Er\"{o}, M.~Flechl, M.~Friedl, R.~Fr\"{u}hwirth\cmsAuthorMark{1}, V.M.~Ghete, C.~Hartl, N.~H\"{o}rmann, J.~Hrubec, M.~Jeitler\cmsAuthorMark{1}, V.~Kn\"{u}nz, A.~K\"{o}nig, M.~Krammer\cmsAuthorMark{1}, I.~Kr\"{a}tschmer, D.~Liko, T.~Matsushita, I.~Mikulec, D.~Rabady\cmsAuthorMark{2}, B.~Rahbaran, H.~Rohringer, J.~Schieck\cmsAuthorMark{1}, R.~Sch\"{o}fbeck, J.~Strauss, W.~Treberer-Treberspurg, W.~Waltenberger, C.-E.~Wulz\cmsAuthorMark{1}
% ...
\vskip\cmsinstskip
% ...
1:~~Also at Vienna University of Technology, Vienna, Austria\\
2:~~Also at CERN, European Organization for Nuclear Research, Geneva, Switzerland\\
% ...

\end{sloppypar}
\end{document}